\documentclass[12pt, epsfig]{article}
\usepackage{amsthm}
\usepackage{amsmath}
\usepackage{amsfonts}
\usepackage{amssymb}
\usepackage{hyperref}
\usepackage{latexsym}
\usepackage{enumerate}
\usepackage{color}
\usepackage{setspace}
\usepackage{comment}
\definecolor{darkgreen}{rgb}{0,0.6,0}
\definecolor{darkblue}{rgb}{0,0,0.3}
\definecolor{darkred}{rgb}{0.7,0,0}

\newcommand{\di}[1]{\partial _{#1}}
\newcommand{\dd}[1]{\nabla _{#1} }

\newcommand{\be}{\begin{equation}}
\newcommand{\ee}{\end{equation}}
\newcommand{\bea}{\begin{eqnarray}}
\newcommand{\eea}{\end{eqnarray}}
\newcommand{\bse}{\begin{subequations}}
\newcommand{\ese}{\end{subequations}}
\newcommand{\beqa}{\begin{eqnarray}}
\newcommand{\eeqa}{\end{eqnarray}}
\newcommand{\beqar}{\begin{eqnarray*}}
\newcommand{\eeqar}{\end{eqnarray*}}
\newcommand{\ba}{\begin{array}}
\newcommand{\ea}{\end{array}}
\newcommand{\bc}{\begin{center}}
\newcommand{\ec}{\end{center}}

\newcommand{\nn}{\nonumber}

\newcommand{\cf}{{\em cf.}\ }

\newcommand{\ads}[1]{AdS$_{#1}$}
\def\sltr{SL$(2,\mathbb{R})$}
\def\sltruon{SL$(2,\mathbb{R})\times$U$(1)^N$}

\newtheorem{lemma}{Lemma}
\def\mydef{\textbf{Definition.} }

\makeatletter \@addtoreset{equation}{section}

\makeatletter\renewcommand\section{\@startsection {section}{1}{\z@}%
                                   {-3.5ex \@plus -1ex \@minus -.2ex}%nn
                                   {2.3ex \@plus.2ex}%
                                   {\normalfont\large\bfseries}}
\renewcommand\subsection{\@startsection{subsection}{2}{\z@}%
                                     {-3.25ex\@plus -1ex \@minus -.2ex}%
                                     {1.5ex \@plus .2ex}%
                                     {\normalfont\bfseries}}

\begin{document}

%--------------------------------------------------------------------------------------------------------------------------------------

\begin{titlepage}
\hfill
\hfill
\vbox{
    \halign{#\hfil         \cr

         IPM/P-2013/033  \cr
         \today\cr }
      }
\vspace*{4mm}

\begin{center}
{\Large{\bf{NHEG Mechanics:\vspace*{2mm}

\centerline{Laws of Near Horizon Extremal Geometry (Thermo)Dynamics\ \ }
}}} \vspace{6mm}

{\large{{\bf K. Hajian$^{\dagger ,\ast}$\footnote{kamalhajian@physics.sharif.edu}, A. Seraj$^\dagger$\footnote{ali\_seraj@ipm.ir}, M. M. Sheikh-Jabbari$^\dagger$\footnote{jabbari@theory.ipm.ac.ir} }}}
\\

\vspace{3mm}

\begin{center}
$^\dagger$\textit{School of Physics, Institute for Research in Fundamental
Sciences (IPM), P.O.Box 19395-5531, Tehran, Iran}\\

$^\ast$\textit{Department of Physics, Sharif University of Technology, \\P. O. Box 11155-9161, Tehran, Iran}
\medskip
\end{center}

\end{center}
\setcounter{footnote}{0}

%------------------------------------------------------------------------------------------------------------------------------------------------------------------------------------------------
\begin{abstract}
\noindent
Near Horizon Extremal Geometries (NHEG) are solutions to gravity theories with SL$(2,\mathbb{R})\times$U$(1)^N$ (for some $N$) symmetry, are smooth geometries and have no event horizon, unlike black holes.
Following the ideas by R. M. Wald,  we derive laws of NHEG dynamics, the analogs of laws of black hole dynamics for the NHEG. Despite the absence of horizon in the NHEG, one may associate an entropy to the NHEG, as a Noether-Wald conserved charge. We work out ``entropy'' and ``entropy perturbation'' laws, which are respectively universal relations between conserved Noether charges corresponding to the NHEG and a system probing the NHEG. Our entropy law is closely related to Sen's entropy function.
We also discuss whether the laws of NHEG dynamics can be obtained from the laws of black hole thermodynamics in the extremal limit.

%The entropy law is closely related to the Sen's entropy function while the entropy variation law may be obtained from appropriate extremal limit of first law of black hole thermodynamics. We comment on the relation our analysis and it relation to Kerr/CFT correspondence.

\bigskip

Keywords:\textit{ Extremal Black Holes, Noether Charges, Entropy Function, Laws of Thermodynamics}

\end{abstract}

\end{titlepage}

%--------------------------------------------------------------------------------------------------------------------------------------
\newpage
\tableofcontents
\setlength{\baselineskip}{1.05 \baselineskip}

\setcounter{footnote}{0}
%------------------------------------------------------------------------------------------------------------------------------------------------------------------------------------------------
%------------------------------------------------------------------------------------------------------------------------------------------------------------------------------------------------
%------------------------------------------------------------------------------------------------------------------------------------------------------------------------------------------------
%------------------------------------------------------------------------------------------------------------------------------------------------------------------------------------------------

%\section{Introduction}
%We might need to have an introduction here.
%------------------------------------------------------------------------------------------------------------------------------------------------------------------------------------------------
%------------------------------------------------------------------------------------------------------------------------------------------------------------------------------------------------
\newpage
\section{Introduction}

Constructing and analyzing solutions to theories of (Einstein) gravity with various kind of matter fields in diverse dimensions has been a very active area of research since the conception of General Relativity. Black holes, stationary solutions with a regular event horizon, has been a class of solutions of particular interest. {We now have classification (not necessarily a complete one) and in some case {uniqueness theorems}  \cite{BH-Uniqueness}  for specific gravity theories. This classification is usually based on the choice of  asymptotic behavior and horizon topology, the charges like mass, angular momenta and electric or magnetic (or possibly dipole) charges and, if there are ``moduli'' in the theory, on the asymptotic values of these moduli scalar fields.}\footnote{This topic started off by notable papers of W. Israel \cite{Israel}, and is more than four decades old, with a rich literature, e.g. see \cite{BH-Uniqueness,Emparan:2008eg} and references therein as some examples.}

Based on the seminal works of Hawking \cite{Hawking} and Bekenstein \cite{Bekenstein}, it was argued that  black holes behave like thermodynamical systems and the four laws of black hole (thermo)dynamics was proposed \cite{BCH}: black hole is a thermodynamical system at the Hawking temperature $T_{H}$ (the temperature of the Hawking radiation as seen by the asymptotic observer) and chemical potentials, the horizon angular velocities $\Omega^i$ and horizon electric/magnetic potentials $\Phi^p$. One can then associate conjugate charges to these, the angular momenta $J_i$, the electric/magentic charges $q_p$ and the (ADM) mass $M$.
These parameters and charges satisfy first law of thermodynamics, if we associate an entropy $S_{BH}$ to the black hole, as Bekenstein and Hawking did; explicitly,\footnote{The moduli (the asymptotic value of scalar fields) may also appear in the first law through a modification of $\delta M$ term. Explicitly, through shifting $\delta M$ to $\delta M- \dfrac{\partial M}{\partial \phi_\alpha} \delta \phi_\alpha$ where $\phi_\alpha$ denotes the moduli \cite{moduli-first-law}.}
\be\label{1st-law}
T_H \delta S_{BH}=\delta M-\sum_i\Omega^{i}\delta J_i-\sum_p\Phi^p\delta q_p\,.
\ee
The remarkable feature of thermodynamical description is its universality, that it is independent of the theory and the specific class of solutions in consideration; it stems from very deep connections between gravity and thermodynamics.

The next conceptual step in the thermodynamical description of black holes appeared in  a series of papers by R. Wald et al. \cite{Wald:1993nt,Iyer:1994ys,Wald-review}. It was argued that not only the charges $J_i$, $q_p$ and $M$, but also the entropy $S_{BH}$ may be viewed as a Noether conserved charge, associated with the  Killing vector field which becomes null (and actually vanishes) at the horizon. Within this approach the first law of black hole thermodynamics was proved. Since our analysis will be based on \cite{Wald:1993nt,Iyer:1994ys}, we will review these works in appendix B.
Among many novel features, Wald's approach clarified  (1) how the charges $J_i$, $q_p$, $M$ and $S_{BH}$ depend on the theory (action), as well as the solution; (2) the significance of gravity equations of motion and dealing with ``solutions'' for having the thermodynamic description (recall that Noether  charges are defined on-shell) and;  (3) the meaning of ``perturbations'' $\delta X$'s appearing in the first law \eqref{1st-law}: The first law is not only about some relations among the parameters defining the class of black hole solutions, the $\delta X$'s are associated with the corresponding charges of a (non-stationary) system probing the black hole background specified by $T_{H}$, $\Omega^i$ and $\Phi^p$; the black hole is seen as a thermodynamical system by the probe.

In search for the micro/statistical mechanical system underlying black holes, the class of extremal black holes, those with $T_H=0$, proved very useful. Extremal black holes may be viewed as the ground state of a system with the same values of $J_i$ and $q_p$ and have generically non-zero entropy, while at zero temperature. It was noted in
\cite{Bardeen-Horowitz,Sen:2005wa,Astefanesei:2006dd} and then rigorously proved in a series of papers \cite{Kunduri:2007vf,Kunduri:2008rs,Kunduri:2013gce} that focusing on a region close to the horizon of extremal black holes we obtain a new class of solutions to the same theory of gravity. This class of solutions, the Near Horizon Extremal Geometries (NHEG's) have the same conserved charges, $J_i$ and $q_p$ as the original black hole, while have no horizon and have a different asymptotic region. As the near horizon limit has been taken, these geometries have no horizon and no singularity. The project of classification and uniqueness theorems for NHEG has been actively pursued in the last decade or so and we have several theorems in four and five dimensions (see \cite{Kunduri:2013gce} for a recent review). We will briefly review these in section \ref{sec-NHEG-review}.

In this work we focus on the NHEG and construct three laws of NHEG (thermo)dynamics. We argue one may associate an entropy to the geometry as the Noether charge associated with a (class of)  Killing vector field(s) which  become null at specific points of spacetime, very similar to what Wald did for black holes. We then work out universal relations among the entropy and other Noether charges of the system. We also work out what resembles first law of (thermo)dynamics for black holes, i.e. a universal relation which governs the relation between perturbations in the entropy and other charges associated with the stationary or non-stationary perturbations of the NHEG.

The rest of this work is organized as follows. In section \ref{sec-NHEG-review}, we review some facts about the NHEG. In section \ref{sec-NHEG-charges}, we compute all Noether charges associated with the symmetries of NHEG. In section \ref{sec-4-entropy-law}, we present the three laws of NHEG mechanics. In section 4.1, we present zeroth law of NHEG mechanics. In section 4.2, work out the ``entropy law'' for the NHEG dynamics, i.e. a universal relation between entropy, which as we argue, itself is a Noether charge,  and other Noether charges of the NHEG. The entropy law formula is closely related to Sen's entropy function \cite{Ent-Funcn-Sen}. In section \ref{sec-5-entropy-variation}, we construct ``entropy perturbation law'' for the NHEG. In section \ref{sec-6-NHEGvs.EBH}, we discuss whether the laws of NHEG dynamics can be constructed from those of black hole dynamics when the black hole becomes an extremal one. We end with discussions and concluding remarks. In the appendices we have gathered some useful relations about the $sl(2,\mathbb{R})$ algebra, a review of Wald-Iyer formulation of the entropy and the first law of black hole thermodynamics, details of the computation of the symplectic form used in section \ref{sec-5-entropy-variation},
and discuss the ``inner-outer horizons permutation symmetry,''  used in section \ref{sec-6-NHEGvs.EBH}.
%\textbf{Note:} In our conventions we take $16\pi G_N=1$.

\section{Near Horizon Extremal Geometries (NHEG)}\label{sec-NHEG-review}

As mentioned in the introduction a generic black hole solution is determined by two class of parameters: those appearing in the thermodynamical description and those associated with the asymptotic values of moduli. There is a largely held idea that all thermodynamical black hole quantities is encoded only in the near horizon data. This viewpoint has been proved for the class of supersymmetric or BPS black holes where it has been shown that the value of the moduli fields at the horizon is independent of their asymptotic values and is completely determined by the (thermodynamical) conserved charges. This observation was called ``attractor mechanism'' \cite{susy-attractor}. It was then realized that \cite{Sen:2005wa,Astefanesei:2006dd,Ent-Funcn-Sen, Sen:2008vm} extremal black holes (which are not necessarily BPS) also exhibit attractor behavior. This means that all the information for ``thermodynamical'' description of black holes\footnote{The term thermodynamical has been put in quotation because extremal black holes are systems at zero temperature and there is really no energy flow. This point will become more clear in the next sections.} is already included in the NHEG. This prompted the study of extremal horizons and exploring the possibility of NHEG uniqueness theorems, which we will review in this section. For further details the reader is referred to the recent comprehensive review \cite{Kunduri:2013gce}.

\subsection{Extremal horizons and near horizon limits}

Extremal black holes are solutions with vanishing surface gravity and hence they do not have a bifurcate horizon. Therefore, it is useful to describe them in a null Gaussian coordinate system \cite{Kunduri:2013gce}:
\be\label{null-Gaussian}
ds^2=2 d\tilde{v}\left(dr + r\tilde{h}_a(r, x)dx^a + \frac12 r^2\tilde{F}(r, x)d\tilde{v}\right)+\tilde\gamma_{ab}(r, x)dx^adx^b\,,
\ee
where the horizon is at $r=0$, and $\tilde\gamma_{ab}$ computed at $r=0$ is the metric on the horizon which is taken to be a smooth, non-degenerate, compact codimension two spacelike surface. One can then readily take the near horizon limit by expanding around $r=0$, setting $r=\epsilon\rho$ and  $v=\tilde{v}/\epsilon$, $\epsilon\to 0$ to obtain
\be\label{null-Gaussian-NH}
ds^2=2 dv\left(d\rho + \rho h_a(x)dx^a + \frac12 \rho^2F(x)d{v}\right)+\gamma_{ab}(x)dx^adx^b\,,
\ee
where $\gamma_{ab}(x)=\tilde\gamma_{ab}(0, x),\ h_a(x)=\tilde{h}_a(0, x),\ F(x)=\tilde{F}(0, x)$. The near-horizon limit has fixed all the $\rho$ dependence. Metric \eqref{null-Gaussian-NH} has translation symmetry along $v$ coordinate, as well as scaling $(v,\rho)\to (v/\lambda, \lambda \rho)$.

Next, one should require \eqref{null-Gaussian-NH} to also satisfy equations of motion. Depending on the theory and its matter content we have some different possibilities for the $h_a$ and $F$ functions and hence the symmetries of the $(v,\rho)$ space. In particular, for ``static'' cases with $dh_a=0$ and when the matter content satisfies strong energy condition the isometry of $(v,\rho)$ part enhances to \sltr. For stationary cases, with four and five dimensional  Einstein-Maxwell-Dilaton (EMD) theory where metric on the space of U(1) gauge fields and dilatons is positive definite (they have non-negative kinetic term) and when the potential of the dilatons is non-positive again we are dealing with a background with \sltruon\ symmetry. Here we do not intend to review in detail the extremal horizon uniqueness theorems. For more detailed and precise discussion see \cite{Kunduri:2013gce}.

As we see for physically interesting cases  the symmetry of the extremal black hole geometry generically enhances to \sltr\ and some other U(1) factors. Therefore, here we only focus on the geometries with such symmetry. Explicitly,\bc
\emph{We define NHEG as the most general geometry with local \sltruon\ symmetry group.}
\ec
Here, we  consider a generic diffeomorphism and gauge invariant theory without specifying the explicit form of the action. (Note that EMD is a special class of such models.)
In general, at most $d-3$ U$(1)$ factors are associated with rotations of the $d$ dimensional spacetime while the rest of them (up to $N$) is the number of gauge fields.

For a generic NHEG we adopt a coordinate system which makes the \sltruon\ symmetry manifest:
\begin{align}\label{NHEG metric}
ds^2=\Gamma\left[-r^2dt^2+\dfrac{dr^2}{r^2}+\sum_{\alpha, \beta=1}^{d-n-2} \Theta_{\alpha \beta}d\theta^\alpha d\theta^\beta+\sum_{i,j=1}^n\gamma_{ij}(d\varphi^i+k^irdt)(d\varphi^j+k^jrdt)\right]\,,
\end{align}
supplemented by a set of gauge fields $A^{(p)}$
\begin{align}\label{eq gauge fields}
A^{(p)}=\sum_{i=1}^n f^{(p)}_i(d\varphi^i+k^irdt)+e^prdt\,.
\end{align}
In the above $i,j=1,\cdots, n$ and $p=n+1,\cdots, N$, and $n\leq d-3$. $\Gamma,\Theta_{\alpha \beta},\gamma_{ij},f^{(p)}_i$ are functions of the polar coordinates $\theta^\alpha$ whose explicit form may be fixed upon imposing equations of motion. $k^i,e^p$ are constants, the constancy of which is a direct consequence of \sltr\ symmetry. A full solution may also involve a number of scalars $\phi_A=\phi_A(\theta^\alpha)$, however, due to the attractor behavior (see \cite{Sen:2005wa} and references therein) the parametric dependence of the scalar fields is completely  fixed by the other charges. So, while these scalars can affect the value of charges, we need not consider them separately in this paper.
We take the constant $r,t$ surfaces, denoted by $H$, to be compact, smooth and non-degenerate. Moreover, we take the metric on $\varphi^i$ space, $\gamma_{ij}$, to be non-degenerate and positive definite.

The geometric part of the \sltruon\ symmetry, which is \sltr$\times$ U$(1)^n$, is generated by the following Killing vector fields (\cf appendix \ref{sec SL2R algebra} for our convention and notations for sl$(2,\mathbb{R})$ algebra.)
\begin{align}\label{sl2r-generators-xi-a}
\xi_1 &=\partial_t\,,\cr
\xi_2 &=t\partial_t-r\partial_r\,,\\
\xi_3 &=\dfrac{1}{2}(t^2+\dfrac{1}{r^2})\partial_t-tr\partial_r -\sum_{i=1}^n\dfrac{k^i}{r}\partial_{\varphi^i}\,,\cr
m_i &=\partial_{\varphi^i}\,,
\end{align}
with the commutation relations:
\begin{align}\label{SL2R}
\left[\xi_1,\xi_2\right]&=\xi_1\,, \quad \quad \left[\xi_2,\xi_3\right]=\xi_3\,,  \quad \quad \left[\xi_1,\xi_3\right]=\xi_2\,,\\
\left[\xi_a,m_i\right]&=0\,,\quad  a\in\{ 1,2,3\} \;\ \text{and},\ \  i\in \{1,\dots , n\}\,.
\end{align}

%------------------------------------------------------------------------------------------------------------------------------------------------------------------------------------------

%------------------------------------------------------------------------------------------------------------------------------------------------------------------------------------------

\subsection{Relation between SL$(2,\mathbb{R})$ and $U(1)$ generators}\label{sec-relation-Killings}
Let us define the \sltr\ vector $n_a,\ a=1,2,3$ as  the unit normal vector to AdS$_2$ in the $\mathbb{R}^{2,1}$ embedding space, i.e. $n_an^a=-1$. In the basis we have used for writing the metric \eqref{NHEG metric} $n_a$ are (see appendix \ref{sec SL2R algebra} for more discussions):
\begin{align}\label{n-a-r-t}
n_1=-r\,,\qquad n_2=-tr\,,\qquad n_3=-\frac{t^2r^2-1}{2r}\,.
\end{align}

Using $n_a$, one has the following relation between the SL$(2,\mathbb{R})$ isometries and U$(1)$ symmetry generators:
\begin{equation}\label{identity xi m}
{n^a\xi_a=k^im_i\,.}
\end{equation}
Note that we have used \sltr\ metric \eqref{Killing form} for raising $a$ index on $n_a$.
To show this recall that the Killing vector $\xi_3$ is
\begin{align}
\xi_3 =\dfrac{1}{2}(t^2+\frac{1}{r^2})\partial_t-tr\partial_r-\sum_i\dfrac{k^i}{r}\partial_{\varphi^i}\,.
\end{align}
Multiplying by $r$ and rewriting the above equation in terms of Killing vectors yields:
\begin{align}
r\xi_3 &=-\dfrac{t^2r^2-1}{2r}\xi_1+tr\xi_2-\sum_ik^i\partial_{\varphi^i}\,,
\end{align}
or
\begin{align}
n_3\xi_1-n_2\xi_2+n_1\xi_3\equiv n^a\xi_a=\sum_i k^i m_i\,.
\end{align}
More detailed analysis and useful identities about the \sltr\ structure is gathered in the appendix \ref{sec SL2R algebra}.

%--------------------------------------------------------------------------------------------------------------------------------------------------------------
\section{NHEG conserved charges}\label{sec-NHEG-charges}

Given a geometry which is (a part of) a solution to a diffeomorphism invariant gravity theory, in the same spirit as the Noether theorem, one may associate a conserved quantity to each Killing vector field. A given solution may also be invariant under some ``internal'' symmetries, like in Maxwell theory, to which one may associate the corresponding Noether charges too. This general argument implies that with the NHEG with \sltruon\ symmetries one can associate $N+3$ conserved Noether charges. In this section we work out those charges.
As reviewed in the appendix B, however, there are always ambiguities in defining Noether charge densities (specially when we are dealing with a symmetry associated with diffeomorphisms). These ambiguities are usually fixed by giving a reference point (e.g. asymptotic ADM charges). Here, we also discuss how those ambiguities may be dealt with in the NHEG case where we do not have a maximally symmetric asymptotic space. Here, following conventions of \cite{Wald:1993nt,Iyer:1994ys}, we use boldface for spacetime forms.

\subsection{Noether charge density of  non-Abelian  symmetries}

Obtaining Noether charge density $\mathbf{Q}$ from the Noether current $\mathbf{J}$ associated to a diffeomorphism generator (\cf appendix \ref{sec symmetry and charges}) is not generally an easy task, but when we are dealing with non-Abelian symmetry groups, this will become straightforward due to construction we discuss below.

Consider  a set of Killing vectors $\xi_a$ which satisfy the following
Lie bracket relations
\begin{align}\label{eq commutations}
[\xi_a,\xi_b]=f_{ab}^{\;\ \;c}\xi_c\,,
\end{align}
where $f_{ab}^{\;\;c}$ are the structural constants of the symmetry Lie algebra $\mathcal{G}$. Let $K_{ab}$ be the metric of the algebra. Then, noting that
\be\label{f-f-K}
f_{ab}^{\;\;\ c}f^{abd}=C_2\ K^{cd}\,,
\ee
where $C_2$ is the second rank Casimir of the algebra in the adjoint representation, we have
\begin{equation}
\xi_a=\frac{1}{C_2} f_a^{\;\ bc}[\xi_b,\xi_c]\,.
\end{equation}
(Note that the indices on the structure constant tensor is raised and lowered by metric $K_{ab}$.)
Next, recalling the definition of the Lie bracket,
\begin{align}
[\xi_b,\xi_c]^\mu &=\xi_b^\nu\nabla_\nu\xi_c^\mu -\xi_c^\nu\nabla_\nu\xi_b^\mu\cr
&=\nabla_\nu\left(\xi_b^{[\nu} \xi_c^{\mu]}\right)\,,
\end{align}
In the second line we have used the Killing property $\dd{\nu}\xi^\nu=0$. Consequently, the Noether current $\mathbf{J}$ (introduced in \eqref{current}) may be written as
\begin{align}\label{eq current}
\mathbf{J}^\mu_{\xi_a}&=\mathbf{\Theta}_{\xi_a}^\mu -\mathcal{L}\xi_a^\mu \cr
&=\frac{2}{C_2}\ \mathcal{L}\ f_a^{\ \ bc} \,\nabla_\nu\left(\xi_b^{\mu} \xi_c^{\nu}\right)\,.
\end{align}
In the second line we have dropped  $\mathbf{\Theta}_{\xi_a}$ term because it is a linear function of $\delta_{\xi_a} \Phi$ and for Killing fields $\delta_\xi \Phi=\mathcal{L}_\xi\Phi=0$. In our notations $\Phi$ stands for all the fields we have in our theory.

One can further simplify \eqref{eq current} using the chain rule and the fact that $\xi_a$'s are isometries of $\mathcal{L}$, i.e. $\xi_a^\nu\nabla_\nu \mathcal{L}=0$, to obtain
\begin{align}
\mathbf{J}^\mu_{\xi_a}&=\nabla_\nu \mathbf{Q}_{\xi_a }^{\mu\nu}\,,
\end{align}
in which
\begin{align}\label{eq J non abelian}
\mathbf{Q}_{\xi_a}^{\mu\nu} =\frac{2}{C_2}\mathcal{L}\ f_a^{\ bc}\xi_b^{\mu} \xi_c^{\nu}\,.
\end{align}

In the presence of (internal) gauge symmetries one should revisit the above analysis: In this case $\delta_\xi\Phi$ is not necessarily zero, $\delta_\xi\Phi$ should be zero up to internal gauge transformations, i.e. generically
\be
\delta_{\xi} \Phi=\delta_{\Lambda}\Phi\,,\qquad \textrm{for some\ } \Lambda=\Lambda(\xi)\,.
\ee
In the diffeomorphism and gauge invariant theories on which we have focused in this work, only the gauge fields $A^{(p)}_\mu$ are subject to the above discussion. So, let us revisit $\mathbf{\Theta}$ term for them:
\begin{align}\label{gauge-field-contrib.}
\mathbf{\Theta}^\mu &=\dfrac{\di{} \mathcal{L}}{\di{}\dd{\mu}A^{(p)}_\nu}\delta A_\nu^{(p)}=\dfrac{\partial \mathcal{L}}{\partial \dd{\mu} A^{(p)}_\nu}\partial _\nu \Lambda^{(p)}\\
&=\dd{\nu}\left(\dfrac{\partial \mathcal{L}}{\partial \dd{\mu} A^{(p)}_\nu}\Lambda ^{(p)}\right) -\Lambda ^{(p)} \dd{\nu} \dfrac{\partial \mathcal{L}}{\partial \dd{\mu}
A^{(p)}_\nu}\,,
\end{align}
where $\Lambda^{(p)}=\Lambda^{(p)}(\xi_a)$ is determined such that $\delta_{\xi_a}A_\mu^{(p)}=\partial_\mu \Lambda^{(p)}$.

Assuming that the action is local and invariant under the gauge $A\rightarrow A+d\Lambda$, it can only be a function  of $F_{\mu\nu}=\partial_{[\mu}A_{\nu]}$ and the second term vanishes due to the field equations for gauge fields in the absence of source\footnote{Note that \sltr\ invariance does not allow for having local sources.}. Therefore,
\begin{align}
\mathbf{\Theta}^\mu =\partial_\nu \mathbf{j}^{\mu\nu},\hspace{1cm}\mathbf{j}^{\mu\nu}=\sum_p \Lambda ^{(p)} \dfrac{\partial \mathcal{L}}{\partial F^{(p)}_{\mu\nu}}\,.
\end{align}
This is the term that should be added to \eqref{eq J non abelian} in the presence of gauge fields and hence the complete form of the Noether charge density  for the generator $\xi_a$ is\footnote{This argument in a straightforward way extends to the non-Abelian internal gauge symmetries and also to  the cases with higher dimensional $p$-forms. Moreover, it is possible that a black hole of non-trivial topology carries a dipole charge while it is neutral, e.g. as in the case of dipole black ring \cite{Emparan-dipole-ring}. These dipole moments do appear in the first law \cite{Copsey-Horowitz} and our analysis may be extended to include these cases.}
\begin{align}
\mathbf{Q}^{\mu\nu}_{\xi_a}= \frac{2}{C_2}\mathcal{L}\ f_a^{\;bc} \xi_b^{\mu} \xi_c^{\nu}+\sum_p \Lambda ^{(p)} \dfrac{\partial \mathcal{L}}{\partial F^{(p)}_{\mu\nu}}\,.
\end{align}

%------------------------------------------------------------------------------------------------------------------------------------------------------------------------------------------------

%------------------------------------------------------------------------------------------------------------------------------------------------------------------------------------------------

\subsection{SL$(2,\mathbb{R})$ conserved charges}\label{sec SL2R charges}

Applying the method of previous subsection, one can compute the conserved charges corresponding to SL$(2,\mathbb{R})$ isometry of NHEG spacetime. It can be seen from \eqref{eq gauge fields} that
\be \label{delta3}
\delta_{\xi_1}A^{(p)}=\delta_{\xi_2}A^{(p)}=0\,,\qquad \delta_{\xi_3}A^{(p)}=-\frac{e^{p}}{r^2}dr=d(\frac{e^{p}}{r})\,,
\ee
and hence $\Lambda^{(p)}_{\xi_3}=\frac{e^{p}}{r}$ (where $\Lambda^{(p)}_{\xi_3}$ is the one appearing in \eqref{gauge-field-contrib.}).

For the sl$(2,\mathbb{R})$ algebra, $C_2=2$ and the Noether charge density for generator $\xi_a$ becomes
\begin{align}\label{eq sl2r Noether potential}
{{\mathbf{Q}_a^{\mu\nu} = \mathcal{L}\ {f_a^{\;bc}}\xi_b^{\mu} \xi_c^{\nu}+\delta_{a3}\sum_p \dfrac{e^p}{r} \dfrac{\partial \mathcal{L}}{\partial F^{(p)}_{\mu\nu}}}}\,.
\end{align}
Using this we can obtain conserved charges corresponding to sl$(2,\mathbb{R})$  Killing vectors by integrating it over the closed surface $H$, which is any of  $(d-2)$-dimensional  $t,r=\text{const}$ surfaces in \eqref{NHEG metric}:
\begin{align}
\mathcal{{Q}}_a\equiv \oint_H d\Sigma_{\mu\nu}\mathbf{Q}_a^{\mu\nu}\,.
\end{align}
Replacing $\mathbf{Q}_a^{\mu\nu}$ from \eqref{eq sl2r Noether potential} and using \eqref{eq identity 3} we obtain
\begin{equation}
\mathcal{Q}_a= \frac{f_a^{\;bc}}{2}\delta_{\xi_b}n_c\oint_H d\Sigma_{tr}\mathcal{L}+\delta_{a3}\sum_p\frac{e^p}{r}  \oint_H d\Sigma_{\mu\nu}\frac{\partial \mathcal{L}}{\partial F^{(p)}_{\mu\nu}}\,,
\end{equation}
where we have used the fact that any function of $r$ can be taken out of the integration, as the integration is on the constant $r$ surface $H$.
Noting  \eqref{eq identity xi n} and recalling the definition of the electric charge
\be
q_p\equiv -\oint_H \mathrm{d} \Sigma_{\mu\nu}\frac{\partial \mathcal{L}}{\partial {F^{^{(p)}}}_{\!\!\mu\nu}}\,,
\ee
we find
\begin{align}\label{Q_i}
{\mathcal{Q}_a=n_a\oint_H d\Sigma_{tr}\mathcal{L}-\delta_{a3}\sum_p\dfrac{e^p}{r}q_p\,.}
\end{align}
It will be more useful to consider the \sltr\ invariant linear combinations of charges $\mathcal{Q}_a$ by multiplying both sides with $n^a$, to obtain
\begin{align}\label{sum Q}
{n^a\mathcal{Q}_a=\sum_p e^pq_p-\oint_H d\Sigma_{tr}\mathcal{L}}\,.
\end{align}

The above analysis, which is based on Noether's theorem, makes it apparent that despite  explicit  $t,r$ dependence, $\mathcal{Q}_a$'s are conserved.  Moreover, in writing \sltr\  charges \eqref{Q_i} we have already fixed the ambiguities associated with Noether-Wald charges discussed in appendix \ref{sec symmetry and charges}. This point will be discussed further in section \ref{sec entropy law}.

%--------------------------------------------------------------------------------------------------------------------------------------------------------------
\subsection{NHEG entropy as a conserved charge}\label{sec entropy Noether}
Despite the fact that the NHEG does not have a (Killing) horizon as  black holes do, recalling that they can be obtained as the near horizon limit of extremal black holes, one may formally associate an entropy to them. To this end, we note that instead of the horizon, the NHEG have  surfaces $H$ ({\it{i.e.}}  surfaces of constant time and radius in the coordinates used to represent the NHEG metric \eqref{NHEG metric}). As discussed in the appendix \ref{sec SL2R algebra}, \sltr\ invariance facilitates defining an (\sltr\ invariant) binormal 2-form (which is dual to the volume form on $H$).
Given these, we can readily write the analogue of Iyer-Wald entropy \cite{Iyer:1994ys} for the NHEG:

\mydef Entropy of the NHEG as a solution of the e.o.m is defined as
\begin{equation}\label{eq NHEG entropy}
\begin{split}
\frac{S}{2 \pi}  &\equiv -\oint_H \text{Vol}(H) \frac{\delta \mathcal{L}}{\delta R_{\mu \nu \alpha \beta}}\epsilon_{\mu \nu}\epsilon_{\alpha \beta}\\
&=-2 \oint_H\mathrm{d} \Sigma_{\mu \nu}{E}^{ \mu \nu \alpha \beta }\epsilon_{\alpha \beta}\,,
\end{split}
\end{equation}
where $H$ is any of the SL$(2,\mathbb{R})$ invariant ($d-$2)-dimensional surfaces, $\epsilon_{\mu\nu}$ is the \sltr\ invariant binormal 2-form, \cf \eqref{def-binormal}, and  $ {E}^{ \mu \nu \alpha \beta } \equiv \frac{\delta \mathcal{L}}{\delta R_{\mu \nu \alpha \beta}}$.

One of the key steps in Wald formulation of ``entropy as a Noether charge'' \cite{Wald:1993nt} is the realization that Killing horizon is associated with a null Killing vector whose dual one-form vanishes on the horizon. In the NHEG we do not have the Killing horizon, however, recalling discussions in section
\ref{sec-relation-Killings}, we indeed have an infinite family of such Killing vector fields:
\begin{equation}\label{eq zeta}
\zeta_H\equiv n^a_H\xi_a-k^im_i\,,
\end{equation}
where $n_H^a$=$n^a(t$=$t_H,r$=$r_H)$ and $n_a$ is given in \eqref{n-a-r-t}. We will prove the following proposition:
\begin{flushleft}
\emph{\hspace{1cm}Conserved charge corresponding to Killing vector $\zeta_H$ is the NHEG Entropy,\\ \hspace{1cm}defined in \eqref{eq NHEG entropy}.}
\end{flushleft}

\begin{proof}
We first note that $\zeta_H$ is  a linear combination of Killing vector fields with constant coefficients ($n^a_H$ and $k^i$ are constants), and hence  $\zeta_H$ is a Killing vector field. Next, we note that according to the proposition 4.1 of the Iyer-Wald paper \cite{Iyer:1994ys} (see appendix \ref{sec symmetry and charges}), the Noether conserved charge corresponding to $\zeta_H$ can be decomposed as
\begin{align}\label{eq  zeta decomp}
\mathcal{Q}_{\zeta _H}=\oint_H d\Sigma_{\mu \nu}\left(W^{\mu \nu}_{\quad \alpha}\zeta_H^\alpha -2\mathbf{E}^{\mu \nu}_{\quad \alpha \beta}\nabla^\alpha \zeta_H^\beta +Y^{\mu \nu} +(dZ)^{\mu\nu}\right)\,,
\end{align}
where $ \mathbf{E}^{ \mu \nu \alpha \beta }= \frac{\delta \mathcal{L}}{\delta R_{\mu \nu \alpha \beta}}$ and $W$ and  $Y$ and $Z$ are covariant quantities which are locally constructed from fields and their derivatives. $Y$ is linear in $\delta _{\zeta_H} \Phi$ and $Z$ is linear in $\zeta_H$ (recall \eqref{n-a-r-t} and \eqref{identity xi m}).
As discussed in the previous section, $\delta_{\zeta_H}\Phi=0$ up to internal gauge transformations. In our case, that is, all $\delta_\xi\Phi=0$, except for $\delta_{\xi_3}A^{(p)}$ which is a pure gauge.
We fix the $Y$ ambiguity requiring physical charges to be gauge independent.
The $W$ and $dZ$ ambiguities are removed, noting that the Killing vector field $\zeta_H$ has been constructed such that  $\zeta_H\vert_{t=t_H,r=r_H}=0$. Therefore,
\begin{align}\label{eq entropy law aux3}
\mathcal{Q}_{\zeta _H}=-2\oint_H d\Sigma_{\mu \nu} \mathbf{E}^{\mu \nu}_{\quad \alpha \beta}\nabla^\alpha \zeta_H^\beta\,.
\end{align}
To determine $\nabla^{\alpha}{\zeta^\beta_H}$, we take covariant derivative of both sides of the identity \eqref{identity xi m},
\begin{align}\label{zeta-H-nabla}
n^a\nabla^\alpha \xi_a^\beta-k^i \nabla^\alpha m_i^\beta=-\xi_a^\beta\nabla^\alpha n^a =\epsilon^{\alpha\beta}\,,
\end{align}
where in the second equation we have \eqref{def-binormal}. The LHS of the above equality may be computed at any $r,t$. In particular, when computed at $r=r_H, t=t_H$ we obtain
\begin{equation}
\nabla^{\alpha}{\zeta^\beta_H}=\epsilon^{\alpha\beta}\,.
\end{equation}
With the above  \eqref{eq entropy law aux3} takes the form
\begin{align}\label{eq entropy Noether}
\mathcal{Q}_{\zeta_H}=-2\oint_H d\Sigma_{\mu \nu}\mathbf{E}^{\mu \nu}_{\quad \alpha \beta}\epsilon^{\alpha\beta}=\frac{S}{2\pi}\,,
\end{align}
which is exactly the NHEG entropy \eqref{eq NHEG entropy} calculated on the surface $H$. It is important to note that although the surface $H$ (defined at constant $t_H,r_H)$ has appeared in the above arguments, the final result is independent of $t_H$ and $r_H$. In other words, there are infinitely many Killing vector fields $\zeta_H$, all leading to the same entropy. This is of course expected because of  the \sltr\ invariance.
\end{proof}
%-------------------------------------------------------------------------------------------------------------------------------------------
%-------------------------------------------------------------------------------------------------------------------------------------------
\section{Laws of NHEG dynamics}\label{sec-4-entropy-law}

%-----------------------------------------------------------------------------------------------------------------------------------------------------------------------------------

%-----------------------------------------------------------------------------------------------------------------------------------------------------------------------------------
{In this section we derive three laws of NHEG mechanics. The first two are describing the NHEG geometry itself, but the third one governs perturbations (or probes) over the NHEG background. The first and third laws resemble the laws of black hole mechanics \cite{BCH}, while ``\textit{entropy law}" has no counterpart for generic  black holes.}

\subsection{Zeroth law of NHEG dynamics}\label{sec-zeroth-law}

Demanding \eqref{NHEG metric} to be \sltr\ invariant, restricts $k^i$ and $e^p$ parameters, while imposing equations of motion will determine other functions there.  In particular, $\xi_3$ is a Killing vector field only if   $\nabla^{\theta^\alpha}\xi_{3}^{\varphi^i}+\nabla^{\varphi^i}\xi_{3}^{\theta^\alpha}\sim \partial _{\theta^\alpha}k^i=0$.
Similarly, if we require that  $ \mathcal{L}_{\xi_3}F^{(p)}=0$, where $F^{(p)}=dA^{(p)}$ and $\mathcal{L}_{\xi_3}$ denotes the Lie derivative w.r.t. the Killing vector $\xi_3$, leads to $\partial _{\theta^\alpha}e^p=0$. That is, $k^i$'s and $e^p$'s should be constants with respect to the coordinates $\theta^\alpha$.

\emph{The constancy of $k^i$ and $e^p$ can be treated as the zeroth law of NHEG dynamics.}

In section \ref{sec-6-NHEGvs.EBH}, we discuss the relation between the NHEG and (near) extremal black holes and show the close connection
between the NHEG zeroth law and the constancy of Hawking temperature and horizon angular velocities. This makes the analogy of NHEG zeroth law and the black hole zeroth law.

\subsection{NHEG entropy law}\label{sec entropy law}
In this section we prove the \emph{``NHEG entropy law''}:
\begin{equation}\label{entropy law}
\boxed{\frac{S}{2\pi}=k^iJ_i+e^pq_p-\oint_H \sqrt{-g}\mathcal{L}\,,}
\end{equation}
where $k^i$ and $e^p$ are constants appearing in the NHEG solution \eqref{NHEG global metric} and \eqref{eq gauge fields}, $J_i$ and $q_p$ denote the corresponding $N$ U(1) charges and
 $\sqrt{-g}=\Gamma^{d/2} \sqrt{\det{\Theta_{\alpha\beta}}\cdot \det\gamma_{ij}}$.
\begin{proof}[Derivation:]
We start by taking covariant derivative from \eqref{eq zeta}
\begin{equation}
-\nabla ^\alpha \zeta_H^\beta=k^i \nabla^\alpha m_i^\beta -n^a_H \nabla^\alpha \xi_a^\beta\,,
\end{equation}
and integrating both sides over $2\oint_H d\Sigma_{\mu \nu}\mathbf{E}^{\mu \nu}_{\quad \alpha \beta}$ :
\begin{equation}\label{eq zeta identity}
-2\oint_H d\Sigma_{\mu \nu}\mathbf{E}^{\mu \nu}_{\quad \alpha \beta}\nabla^\alpha \zeta_H^\beta=2\oint_H d\Sigma_{\mu \nu}\mathbf{E}^{\mu \nu}_{\quad \alpha \beta}\big(k^i \nabla^\alpha m_i^\beta -n^a_H \nabla^\alpha \xi_a^\beta \big)\,.
\end{equation}

Next, we note that as discussed in the appendix \ref{sec symmetry and charges}, there is a Noether conserved charge associated each of the  Killing vector fields $\zeta_H$, $\xi_a$ and $m_i$, but these conserved charges come with three kind of $W, Y, dZ$ ambiguities
\begin{align}%\label{eq decomposition all}
\mathcal{Q}_{\zeta_H}&=\oint_H d\Sigma_{\mu \nu}\lbrack W^{\mu \nu}_{ \,\,\,\alpha}{\zeta_H}^\alpha -2\mathbf{E}^{\mu \nu}_{\quad \alpha \beta}\nabla^\alpha {\zeta_H}^\beta +Y_{\zeta_H}^{\mu\nu}+(dZ_{\zeta_H})^{\mu\nu}\rbrack\,,\nonumber\\
\mathcal{Q}_{m_i}&=\oint_H d\Sigma_{\mu \nu}\lbrack W^{\mu \nu}_{ \,\,\,\alpha}m_i^\alpha -2\mathbf{E}^{\mu \nu}_{\quad \alpha \beta}\nabla^\alpha m_i^\beta +Y_{m_i}^{\mu\nu}+(dZ_{m_i})^{\mu\nu}\rbrack\,, \cr
\mathcal{Q}_{\xi_a}&=\oint_H d\Sigma_{\mu \nu}\lbrack W^{\mu \nu}_{ \,\,\,\alpha}\xi_a^\alpha -2\mathbf{E}^{\mu \nu}_{\quad \alpha \beta}\nabla^\alpha \xi_a^\beta +Y_{\xi_a}^{\mu\nu}+(dZ_{\xi_a})^{\mu\nu}\rbrack\,.\nonumber
\end{align}
Computed ``at the horizon'' where $\zeta_H$ is zero, the $W$ and $dZ$ terms in $\mathcal{Q}_{\zeta_H}$ vanish. Similarly, in the following linear combination of other charges
$$
\sum_a n_H^a \mathcal{Q}_{\xi_a}-\sum_i k^i\mathcal{Q}_{m_i}\,,
$$
the $W$ and $dZ$ terms also vanish. Therefore, \eqref{eq zeta identity} becomes
$$
\mathcal{Q}_{\zeta_H}-(\sum_a n_H^a \mathcal{Q}_{\xi_a}-\sum_i k^i\mathcal{Q}_{m_i})=
\oint_H d\Sigma_{\mu \nu}\left(Y_{\zeta_H}^{\mu\nu}- n_H^a Y_{\xi_a}^{\mu\nu}+k^iY_{m_i}^{\mu\nu} \right)\,.
$$
The RHS of the above equation is zero because $\delta_\xi\Phi$ is linear in $\xi$ (or in $\nabla\xi$) as well as in $\Phi$ (or in $\nabla\Phi$), and hence
$\delta_{\zeta_H}\Phi-(n^a_H\delta_{\xi_a}\Phi-k^i\delta_{m_i}\Phi)=\delta_{\zeta_H-n_H^a\xi_a+k^im_i}\Phi=0$. In summary, all the three $W$, $Y$ and $dZ$ type ambiguities cancel out from the two sides of the equality and we obtain
\be\label{eq entropy law aux4}
\mathcal{Q}_{\zeta_H}=\sum_a n_H^a \mathcal{Q}_{\xi_a}-\sum_i k^i\mathcal{Q}_{m_i}\,.
\ee
With a similar reasoning one can show that the above equation holds when we replace $\mathcal{Q}_{\zeta_H}$ by $S/(2\pi)$ (\cf \eqref{eq entropy Noether}),
$\mathcal{Q}_{m_i}$ by physical angular momenta $J_i$, and $n_H^a \mathcal{Q}_{\xi_a}$ from \eqref{sum Q}. We hence obtain the desired entropy law expression
\eqref{entropy law}.
\end{proof}

Before closing this section some comments are in order:
\begin{enumerate}
\item Eq.\eqref{entropy law} is universal, meaning that it is the relation between conserved charges associated with any NHEG solution to any diffeomorphism invariant theory (of gravity).
\item In the above we have used the fact that the LHS of \eqref{sum Q} is \sltr\ invariant and hence can be computed at any arbitrary constant $t,r$ surface.
\item The entropy law \eqref{entropy law} is  a manifestation of the fact that the \sltr\ and U$(1)$ generators mix with each other, as is manifest, e.g. from \eqref{sl2r-generators-xi-a}. Explicitly, the $\xi_3$ Killing vector also involves a $k^i\partial_{\phi^i}$ term \eqref{sl2r-generators-xi-a}.
     
\item The entropy law (and also the entropy perturbation law \eqref{entrop-pert-law}) are invariant under permutation of $N$ $U(1)$ symmetries.

\item We stress that such a universal relation between entropy and other thermodynamical quantities/conserved charges does not exist for generic black holes. As we will discuss further in following sections, the ``first law'' of black hole thermodynamics deals with perturbations of these parameters and not themselves. Note also that Smarr-like formulas which may resemble our entropy law, are not universal and are solution and/or theory dependent.
\item The reason why our derivation of entropy law (or in other words, Wald's derivation) does not hold for generic black holes is presence of ambiguities
we discussed in some detail, and in particular the fact that these ambiguities should be computed and compared at different locations in the black hole geometry. In our case, unlike the black hole case, we have vanishing Killing vector $\zeta_H$ for any $t_H,r_H$. We will elaborate on this point further in the next sections.
\item Our derivation is based on Noether conserved charges and hence makes clear the role of being on-shell. In particular, in the last term in \eqref{entropy law}, the Lagrangian $\mathcal{L}$ should be computed on the NHEG solution.
\item The entropy \eqref{eq entropy Noether} is a conserved charge associated with a vanishing Killing vector field $\zeta_H$, although NHEG does not have a horizon. The entropy is completely determined by the geometry and not other fields, although other fields affect the geometry through Einstein equations.
\item Note that in our ansatz for gauge fields \eqref{eq gauge fields} we have already included possibility of having a non-zero magnetic flux (through the $f^{(p)}_i d\varphi^i$ term). As expected, the magnetic and electric flux (denoted through $e^p$) appear asymmetrically in our entropy law; magnetic flux appears only through the Lagrangian term.
\item In our derivation it is clear that the terms in the RHS of the entropy law are associated with $N$ U$(1)$ symmetries of the system and the corresponding conserved charges. The dilaton-type scalar fields (or moduli) which are not associated with any symmetry can only appear through the Lagrangian term. This is a realization of the attractor behavior \cite{Sen:2005wa,Astefanesei:2006dd,Sen:2008vm} in our setup.
\item Our entropy law is closely related to Sen's entropy function \cite{Sen:2005wa,Ent-Funcn-Sen}.\footnote{We point out that in the entropy function formulation one is prescribed to start from an ``off-shell entropy functional'' defined on the NHEG solution \eqref{NHEG metric} and \eqref{eq gauge fields}, and then find equations of motion by setting zero variations of this entropy functional with respect to unknown functions or parameters of the NHEG solution ansatz. Computing the value of this entropy functional on the solutions to these equations of motion is shown to reproduce Wald entropy for extremal black holes \cite{Sen:2005wa,Sen:2008vm}.} However, our derivation is quite different; specifically we note that our derivation is completely based on the NHEG and not the extremal black hole. Therefore, we need not deal with the issues which may arise in the usage of Wald entropy formula which is derived for bifurcate horizons, for extremal horizons. Further discussion related to this point can be found in section \ref{sec-6-NHEGvs.EBH}.
\end{enumerate}

%-------------------------------------------------------------------------------------------------------------------------------------------
%------------------------------------------------------------------------------------------------------------------------------------------------------------------------------------------------
%------------------------------------------------------------------------------------------------------------------------------------------------------------------------------------------------
\subsection{NHEG entropy perturbation law}\label{sec-5-entropy-variation}

In the previous section we derived the NHEG entropy law, which is a relation among conserved Noether-Wald charges of the NHEG which is a solution to equations of motion for a given gravity theory with our desired \sltruon\ symmetry. As pointed out this relation has no universal analog for generic black holes. In this section we construct  the analog of the first law of thermodynamics for the NHEG.

{To this end, let us denote the NHEG solution by the field configuration $\Phi_0$ and consider a perturbation around it $\delta\Phi$. The configuration $\Phi_0+\delta\Phi$ is not necessarily of the form of NHEG, however, we assume that the perturbations $\delta\Phi$ satisfy linearized equations of motion around the NHEG background solution $\Phi_0$. Therefore, $\delta\Phi$  can also be labeled by the same charges as the background. Let us denote these charges by $\delta J_i$, $\delta q_p$ and $\delta S$. Our discussions here are basically paralleling those in \cite{Iyer:1994ys} for ordinary black hole. However, as we will see below, the case of NHEG has its own specific and novel features. Under specific conditions over field perturbations $\delta\Phi$ which are listed in the end of this section,  we prove the ``\textit{entropy perturbation law}'' relating different charges of the probe:}
\begin{equation}\label{entrop-pert-law}
{\boxed{\frac{\delta S}{2\pi}=k^i \delta J_i +e^p\delta q_p\,}}
\end{equation}

\begin{proof}[Derivation:] Noether current corresponding to  the diffeomorphism generated by $\zeta_H$ is (see appendix \ref{sec symmetry and charges} for notations):
\begin{equation}\label{eq standard Noether current}
\mathbf{J}_{\zeta_H}=\mathbf{\Theta} (\Phi,\delta_{\zeta_H} \Phi)-\zeta_H \! \cdot \! \mathcal{\mathbf{L}}\,,
\end{equation}
where $\zeta_H$ is the Killing vector field defined in \eqref{eq zeta}. We will use $\xi\cdot \mathbf{X}$ to denote the contraction of the vector $\xi$ with the first index of the form $\mathbf{X}$, which is usually written as $i_\xi \mathbf{X}$.  Let us now consider variations in \eqref{eq standard Noether current} associated with $\Phi_0\to \Phi_0+\delta\Phi$:
\begin{equation} \label{eq xi L0}
\delta \mathbf{J}_{\zeta_H}=\delta \lbrack \mathbf{\Theta} (\Phi,\delta_{\zeta_H} \Phi)\rbrack -{\zeta_H} \! \cdot \! \delta \mathbf{L}\,.
\end{equation}
We assume that the variations do not alter the quantities attributed to the background. In particular, this means that $\delta\zeta_H, \delta\xi_a, \delta m_i$ are all vanishing (as they do in the case of black holes \cite{Wald:1993nt,Iyer:1994ys}). In this sense these variations are considered as perturbations or \textit{probes} over the NHEG. Let us start our analysis from the  last term in \eqref{eq xi L0}:
\begin{equation}
\delta \mathbf{L}= \mathbf{E}_i \delta \Phi ^i+\mathrm{d} \mathbf{\Theta}(\Phi_0 , \delta \Phi)\,.
\end{equation}
The first term vanishes due to the on-shell condition  and the second term is simplified recalling the identity  ${\xi} \! \cdot    \mathrm{d} \mathbf{\Theta} =\delta_{\xi} \mathbf{\Theta}-\mathrm{d}({\xi} \!\cdot\! \mathbf{\Theta})$ which is valid for any diffeomorphism $\xi$, therefore,
\begin{equation}\label{eq xi L}
{\zeta_H} \! \cdot \! \delta \mathbf{L}=\delta_{\zeta_H} \mathbf{\Theta}(\Phi_0,\delta \Phi) - \mathrm{d} ({\zeta_H} \! \cdot \! \mathbf{\Theta} (\Phi_0,\delta \Phi))\,.
\end{equation}
Inserting the above into \eqref{eq xi L0} we obtain
\begin{equation}\label{deltaJ}
\delta \mathbf{J}_{\zeta_H}=\boldsymbol{\omega}(\Phi_0,\delta \Phi,\delta_{\zeta_H}\Phi)  + \mathrm{d} ({\zeta_H} \! \cdot \! \mathbf{\Theta} (\Phi_0,\delta \Phi))\,.
\end{equation}
where
\begin{align}
\boldsymbol{\omega}(\Phi_0,\delta_1 \Phi,\delta_2\Phi)  \equiv\delta_1 \mathbf{ \Theta} (\Phi_0,\delta_2 \Phi) -\delta_2 \mathbf{\Theta} (\Phi_0,\delta_1 \Phi)
\end{align}
is the \textit{symplectic current}, the $(d-1)$-form associated with variations $\delta_1,\delta_2$, and is bilinear in its arguments \cite{Wald:1993nt}. This implies that for Killing vectors $\xi$ with $\delta_\xi\Phi_0=0$, the symplectic form vanishes. However, in presence of gauge fields $\delta_\xi\Phi_0$ need not vanish for a symmetry, it may be non-zero up to gauge transformations. In particular, as we have already seen in previous section, this is the case for the third Killing vector $\xi_3$ and the corresponding  symplectic current $\boldsymbol{\omega}(\Phi_0,\delta \Phi,\delta_{\xi_3}\Phi)$ does not vanish. This feature (which was not relevant for the discussions of black holes \cite{Wald:1993nt,Iyer:1994ys}) has an important role in our derivation of the entropy perturbation law.

%
%\begin{equation}\label{deltaJ}
%\delta \mathbf{J}_{\zeta_H}= \mathrm{d} ({\zeta_H} \! \cdot \! \mathbf{\Theta} (\Phi_0,\delta \Phi))\,.
%\end{equation}
The current $\mathbf{J}_{\zeta_H}$ is conserved \emph{on-shell}, i.e $\mathrm{d}\mathbf{J}_{\zeta_H}=0$, so one can associate a conserved charge $d-2$ form $\mathbf{Q}_{\zeta_H}$, $\mathbf{J}_{\zeta_H}=\mathrm{d} \mathbf{Q}_{\zeta_H}$,  to the symmetry generated by $\zeta_H$. Moreover, when the solution is deformed by a perturbation which is a solution to the linearized equations of motion, the relation $\mathrm{d}\mathbf{J}_{\zeta_H}=0$ still holds even if the perturbation is not symmetric under $\zeta_H$ (i.e. $\delta_{\zeta_H}(\delta\Phi)\neq 0$). In other words, one can take the variation of the relation $\mathbf{J}_{\zeta_H}=\mathrm{d} \mathbf{Q}_{\zeta_H}$ and arrive at \cite{Wald:1993nt}
\begin{align}\label{delta,d}
\delta \mathbf{J}_{\zeta_H}=\delta \mathrm{d} \mathbf{Q}_{\zeta_H}=\mathrm{d} \delta \mathbf{Q}_{\zeta_H}\,.
\end{align}
From the above equation, we also learn that perturbations over a background can be labeled by the charges corresponding to the background symmetries, although they do not carry those symmetries. Using \eqref{delta,d} in \eqref{deltaJ} yields
\begin{equation}\label{conservation-zetaH}
\boldsymbol{\omega}(\Phi_0,\delta \Phi,\delta_{\zeta_H}\Phi)  =\mathrm{d}\Big(\delta  \mathbf{Q}_{\zeta_H}- {\zeta_H} \! \cdot \! \mathbf{\Theta} (\Phi_0,\delta \Phi)\Big)\,.
\end{equation}

We integrate the above ``conservation equation'' over a timelike hypersurface $\Sigma$ bounded between two radii  $r=r_H,\;r=\infty$. The hypersurface $\Sigma$ can be simply chosen as a constant time surface $t=t_H$. The interior boundary $r=r_H$ is necessary, since AdS$_2$ does not have a compact interior. As discussed before, the surface $H$ will play the role of horizon on which we define the entropy of NHEG. The $r=\infty$ choice for the other boundary, is a convenient choice because the extra terms appearing due to gauge transformations vanish (\emph{cf.} appendix \ref{appendix symplectic form}, and in particular discussions around \eqref{Lambda-infinity}).
Following \cite{Wald:1993nt}, we define the \textit{symplectic form} associated with $\Sigma$ as
\begin{align}
\Omega(\Phi_0,\delta_1 \Phi,\delta_2\Phi) \equiv \int_\Sigma\boldsymbol{\omega}(\Phi_0,\delta_1 \Phi,\delta_2\Phi)\,.
\end{align}
Integrating \eqref{conservation-zetaH} over $\Sigma$ then yields:
\begin{align}\label{omega vs Q}
\nonumber\Omega(\Phi_0,\delta \Phi,\delta_{\zeta_H}\Phi) &= \oint_{\partial\Sigma}\Big(\delta  \mathbf{Q}_{\zeta_H}- {\zeta_H} \! \cdot \! \mathbf{\Theta} (\Phi_0,\delta \Phi)\Big)\\
&=\oint_{\infty}\Big(\delta  \mathbf{Q}_{\zeta_H}- {\zeta_H} \! \cdot \! \mathbf{\Theta} (\Phi_0,\delta \Phi)\Big)-\oint_{H}\delta  \mathbf{Q}_{\zeta_H}
\end{align}
where in the first line we have used the Stokes theorem to convert the integral over $\Sigma$ to an integral over its boundary $\partial\Sigma$ and in the second line, we used the fact that $\zeta_H=n_H^a \xi_a-k^im_i$ vanishes on $H$. Since the charge perturbation $\delta\mathbf{Q}_{\zeta_H}$ is linear in the vector $\zeta_H$, one can expand the first term on RHS of \eqref{omega vs Q}
\begin{align}
\Omega(\Phi_0,\delta \Phi,\delta_{\zeta_H}\Phi) &= n_H^a\oint_{\infty}\Big(\delta  \mathbf{Q}_{a}- {\xi_a} \! \cdot \! \mathbf{\Theta}\Big)-k^i\oint_{\infty}\Big(\delta  \mathbf{Q}_{m_i}- {m_i} \! \cdot \! \mathbf{\Theta}\Big)-\oint_{H}\delta  \mathbf{Q}_{\zeta_H}\,.
\end{align}
$m_i$ is tangent to the boundary surface and hence the pullback of ${m_i} \! \cdot \! \mathbf{\Theta}$ over the surface $r=\infty$ vanishes, and we have
\begin{align}\label{Omega}
\Omega(\Phi_0,\delta \Phi,\delta_{\zeta_H}\Phi) &= n_H^a\delta \mathcal{E}_a-k^i\oint_{\infty}\delta  \mathbf{Q}_{m_i}-\oint_{H}\delta  \mathbf{Q}_{\zeta_H}\,,
\end{align}
where
\begin{align}\label{symplectic current}
\delta \mathcal{E}_a\equiv\oint_\infty\ (\delta \mathbf{Q}_{\xi_a} -{\xi_a} .\mathbf{\Theta})\,,
\end{align}
is the canonical generator of the symmetry $\xi_a$ in the covariant phase space \cite{oai:arXiv.org:gr-qc/9503052}.

{The technical details of computation of $\Omega(\Phi_0,\delta \Phi,\delta_{\zeta_H}\Phi)$ is given in the appendix \ref{appendix symplectic form}, where it is shown that  $$\Omega(\Phi_0,\delta \Phi,\delta_{\zeta_H}\Phi) =-e^p \delta q_p.$$ }Substituting this result into \eqref{Omega} yields
\begin{align}\label{delta Q zeta}
\oint_{H}\delta \mathbf{Q}_{\zeta_H}&=k^i\delta J_i+e^{p}\delta q_{p}+n_H^a\delta \mathcal{E}_a\,,
\end{align}
where $\delta J_i$ is the angular momentum corresponding to the rotational symmetry $m_i$
\begin{align}
\delta J_i\equiv-\oint_\infty \delta \mathbf{Q}_{m_i}\,.
\end{align}
(Since pullback of ${m_i} \! \cdot \! \mathbf{\Theta}$ vanishes over any constant $t,r$ surface on NHEG, one can show that in the above equation $\delta J_i$  
could be computed with the integral at $\infty$ replaced by any $r=r_H$ surface.)

To show that the left side of \eqref{delta Q zeta}  is actually the perturbation of entropy $\delta S$, we should discuss ambiguities of $\delta \mathbf{Q}_{\zeta_H}$. Any Noether charge can be decomposed as in \eqref{eq  zeta decomp} with $W$, $Y$ and $dZ$ ambiguities. The $W$ and $dZ$ ambiguities vanish since they are linear in $\zeta_H$, which vanishes at surface $H$. The $\delta Y$ ambiguity, which is proportional to variation of fields $\delta_\xi\Phi$ needs more attention. Since $\zeta_H=0$,  at surface $H$, $\delta_{\zeta_H}\Phi=0$. This implies that Y vanishes on background over $H$, and also that its perturbation is given by
\begin{align}\label{eq deltaY}
\delta Y(\Phi_0,\delta_{\zeta_H}\Phi)&=Y(\Phi_0,\delta\delta_{\zeta_H} \Phi)\nonumber\\
&=Y(\Phi_0,\delta_{\zeta_H}\delta \Phi)\nonumber\\
&=\delta_{\zeta_H}Y(\Phi_0 ,\delta \Phi)\nonumber\\
&=\zeta_H \cdot dY+d(Y\cdot \zeta_H)\,.
\end{align}
In the above we have used the fact that since $\delta\zeta_H=0$, we can  interchange $\delta_{\zeta_H}$ and $\delta$. Equation \eqref{eq deltaY} is linear in the generator $\zeta_H$, does not contribute to the left hand side of \eqref{delta Q zeta} and therefore
\begin{equation}
\delta \oint _H   \mathbf{Q}_{\zeta_H}=-2\delta\oint_H d\Sigma_{\mu \nu}\mathbf{E}^{\mu \nu}_{\quad \alpha \beta}\nabla^\alpha \zeta_H^\beta =\frac{\delta S}{2\pi}\,.
\end{equation}
so
\begin{align}\label{entropy-pert-Ea}
\frac{\delta S}{2\pi}&=k^i\delta J_i+e^{p}\delta q_{p}+n_H^a\delta \mathcal{E}_a\,.
\end{align}
Analysis of \cite{Amsel:2009ev} indicates that the NHEG background is stable for a class of field perturbation which satisfy  certain boundary conditions. As we will show in our upcoming work \cite{Progress}, this stability  condition implies $\delta \mathcal{E}_a=0$. Dropping the last term in \eqref{entropy-pert-Ea} by the choice of boundary conditions, we arrive at the desired entropy perturbation law \eqref{entrop-pert-law}.
\end{proof}
{To end this section we summarize the assumptions over the field perturbations which resulted in the entropy perturbation law \eqref{entrop-pert-law}:
\begin{itemize}
\item Perturbations should satisfy the linearized field equations.
\item Perturbations are restricted to those for which  \sltr  $\;$ charges vanish, i.e  $\delta \mathcal{E}_a=0$. This is typically done by choosing a set of boundary conditions.
\end{itemize}
We also note that the variation $\delta$ does not affect the Killing vectors associated with the background, i.e $\delta\zeta_H= \delta\xi_a= \delta m_i=0$.
}
\section{NHEG vs. extremal black hole}\label{sec-6-NHEGvs.EBH}

So far we focused on NHEG as an interesting class of solutions to gravity theories and introduced and worked out three laws of NHEG dynamics. NHEG, as the name implies, is related to extremal black holes and one may wonder if laws of NHEG dynamics can be (directly) related to the laws of extremal black hole thermodynamics. This question has of course been discussed and studied in the literature from various different perspectives, see in particular \cite{First-law, Compere}. This section is mainly meant to fill some gaps remaining in the literature about the connection of NHEG and extremal black holes.

The most general form of the metric of a stationary and axisymmetric black hole possessing some $U(1)$ gauge fields, can be written in the ADM form as
\be\begin{split}\label{BH geometry}
ds^2&=-f d\tau^2+g_{\rho\rho}d\rho^2+\tilde g_{\alpha\beta}d\theta^\alpha d\theta^\beta +g_{ij}(d\psi^i-\omega^i d\tau)(d\psi^j-\omega^j d\tau)\,,\cr
\tilde{A}^{(p)}&=\Phi^{(p)}\: d\tau +\sum_i \mu^{(p)}_i (d\psi^i-\omega^i d\tau)\,,
\end{split}\ee
where $f, g_{\rho\rho}, \tilde g_{\alpha\beta}, g_{ij}, \omega^i$ and $\Phi^{(p)}, \mu^{(p)}_i$ are functions of $\rho,\theta^\alpha$ and $i,j=1,2,\cdots, n$ and $p=n+1,\cdots, N$. The horizons of black hole are at the roots of $g^{\rho\rho}$,
\begin{align}\label{Delta}
g_{\rho\rho}=\dfrac{1}{D^2(\rho,\theta^\alpha)\Delta(\rho)}\,,\qquad \Delta=\prod_m (\rho-r_m)\,,
\end{align}
where we assume the function $D$ to be analytic and nonvanishing everywhere. Due to the smoothness of metric on the horizons $f$ can always be written in the following form:
\begin{align}
f=C^2(\rho,\theta)\Delta(\rho)\,.
\end{align}
In four dimensions the black hole has at most two horizons (e.g. see \cite{BH-Uniqueness}) and $\Delta=(\rho-r_+)(\rho-r_-)$. When there exist more than two horizons, we call the  outermost two horizons as $r_- , r_+\; (r_+>r_-)$. The constants $r_+,r_-$ are two parameters characterizing the black hole. We introduce $r_h,\epsilon$ instead of $r_\pm$ as:
\begin{align}\label{r_h}
r_h\equiv (r_++r_-)/2\,, \qquad
\epsilon \equiv (r_+-r_-)/2\,.
\end{align}
The above notation turns out to be useful since $\epsilon$ is a good measure of black hole temperature $T_H$. Hawking temperature of the black hole can be found requiring the near horizon metric in the Euclidean sector to be free of conical singularity  (e.g see \cite{Solodukhin:1994yz}), leading to \cite{SheikhJabbaria:2011gc}
\begin{align}\label{temperature}
 T_H =\dfrac{1}{2\pi}\left.\sqrt{g^{\rho\rho}}\:\partial_\rho \sqrt{f}\right\vert_{\rho=r_+}=\dfrac{CD}{4\pi}(r_+-r_-)=\dfrac{CD}{2\pi}\epsilon\,,
\end{align}
where in the above $C$ and $D$ are computed at the horizon $\rho=r_+$.
Constancy of Hawking temperature on the horizon implies that $C(r_+,\theta)D(r_+,\theta)$ is a constant on the horizon \cite{SheikhJabbaria:2011gc}.
In the extremal limit, $\epsilon\rightarrow 0$ and $\Delta$ in \eqref{Delta} will have a double root at $\rho=r_e$.
%\subsection{Near Extremal Spinning Black holes}
\subsection{Near horizon limit of extremal black holes}\label{sec6.1-NH-limit}
From now on we will focus on the extremal case, $r_+=r_-=r_e$. To take the near horizon limit let us first make the coordinate and gauge transformations
\begin{align}
\rho &=r_e(1+\lambda r)\,,\qquad
\tau =\dfrac{\alpha r_e t}{\lambda}\\
\varphi^i &=\psi^i-\Omega^i \tau\,,\qquad A^{(p)}=\tilde{A}^{(p)}+d\Lambda, \hspace{1cm}\Lambda= -\Phi^{(p)}\vert_{r_e}\tau
\end{align}
where $\Omega^i=\omega^i(r_e)$ is the horizon angular velocity and $\Phi^{(p)}\vert_{r_e}$ is the horizon electric potential.
In the first line we scale $\rho-r_e$ and $\tau$ inversely by a factor $\lambda$ and $\alpha$ is a suitable constant to get the most simple form for the near horizon metric. $\lambda$ is the parameter which we send to zero once we take the limit. The shift in $\psi^i$ takes us to the frame co-rotating with the black hole. In the last equation, we have used the gauge symmetry  in order to remove the infinities resulting from the limit $\lambda\rightarrow 0$. Upon these transformations the near horizon geometry (obtained in the $\lambda\to 0$ limit) becomes
\begin{align}
ds^2
%&=-C^2(\rho-r_h)^2 d\tau^2+\dfrac{d\rho^2}{D^2(\rho-r_h)^2}+g_{\alpha\beta}d\theta^\alpha d\theta^\beta+g_{ij}(d\psi^i-\omega^i d\tau)(d\psi^j-\omega^j %d\tau)\\
=\dfrac{1}{D^2}\left[- r^2dt^2+\dfrac{dr^2}{r^2}+D^2\tilde g_{\alpha\beta}d\theta^\alpha d\theta^\beta
+D^2g_{ij}(d\varphi^i+(\Omega^i-\omega^i) d\tau)(d\varphi^j+(\Omega^j-\omega^j) d\tau)\right]\,,
\end{align}
where we used the fact that $CD=const$ on the horizon and chose
\begin{align}
\alpha r_e^2=\dfrac{1}{CD}\,.
\end{align}
Recalling that $\Omega^i=\omega^i\vert_{r_e}$, we arrive at the general form:
\begin{align}\label{NH-geometry}
ds^2&=\Gamma\left[-r^2dt^2+\dfrac{dr^2}{r^2}+g_{\alpha\beta}d\theta^\alpha d\theta^\beta
+\gamma_{ij}(d\varphi^i +k^irdt)(d\varphi^j +k^j rdt)\right]\\
A^{(p)}&=e^{(p)}rdt+\sum_i \mu^{(p)}_i(d\varphi^i+k^i r dt)\,,
\end{align}
in which
\begin{align}\label{parameters}
 \Gamma =\dfrac{1}{D^{2}}\bigg|_{\rho=r_e},\quad\gamma_{ij} =D^2 g_{ij}\bigg|_{\rho=r_e}\,,\quad
k^i =-\dfrac{1}{CD}\dfrac{\partial \omega^i}{\partial \rho}\bigg|_{\rho=r_e}\,,\quad e^{(p)}=\dfrac{1}{CD}\dfrac{\partial \Phi^{(p)}}{\partial \rho}\bigg|_{\rho=r_e}\,.
\end{align}
The above is, as expected, the same as the NHEG ansatz \eqref{NHEG metric} and \eqref{eq gauge fields}.

We first show that smoothness of black hole geometry \eqref{BH geometry} forces $\partial_\rho\omega^i$ to be constant on the horizon, and $k^i$ are hence constants in the NHEG. A more detailed proof for this has appeared in \cite{smoothness-2008} (see the appendix there). However, here we give an alternative argument. Analysis of finiteness of curvature invariants for solutions to field equations of the form \eqref{NH-geometry}
reveals that  $(\partial_{\theta^{\alpha}} \omega^i)^2\sim (\rho-r_e)^{2\alpha}$, with $\alpha>1$.  Therefore, $\partial_\rho \partial_{\theta^\alpha}\omega^i\big|_{\rho=r_e}= \partial_{\theta^\alpha}\partial_\rho\omega^i\big|_{\rho=r_e}=0$. So, not only $\partial_{\theta^\alpha}\omega^i=0$ on the horizon (which means that angular velocity is constant on the horizon), but also $\partial_\rho\partial_{\theta^\alpha} \omega^i=0$ which means that $ \partial_\rho\omega^i$ is constant at the horizon of extremal black holes. Using the third equation of \eqref{parameters}, we find that $k^i$ are $\theta$ independent and hence constants. This is a restatement of the zeroth law for NHEG geometries (\cf section \ref{sec-zeroth-law}).

\subsection{NHEG entropy perturbation law and near horizon limit}
Here we briefly review what was done in \cite{First-law} (see also \cite{Compere,Compere-higher-derivative}): One can indeed derive ``entropy variation law'' of NHEG from taking the extremal limit, starting from first law of thermodynamics for \emph{near extremal} black holes. To this end, we recall
the first law of black holes stating how perturbation of entropy is related to the perturbations of mass and other conserved charges of any black hole:
\begin{align}\label{1st law}
\delta M=T_H\delta S+\sum_i\Omega^i \delta J_i +\sum_p \Phi^p \delta q_p\,.
\end{align}
At the extremal point where $T_H=0$ the above reduces to $\delta M=\sum_i\Omega^i \delta J_i +\sum_p \Phi^p \delta q_p$, which may in principle be integrated to get the BPS relation $M=M(J_i,q_p)$. In the near extremal case when $T_H\sim \epsilon$, one may then make a low temperature expansion of all thermodynamics quantities  in powers of $\epsilon$. For black holes, we have the crucial relation that \cite{First-law} $\delta M-\Omega_{ext}^i\delta J_i-\Phi_{ext}^p\delta q_p\sim \epsilon^2$, and hence to the leading order in $\epsilon$ the first law reduces to
\begin{align}\label{limiting-first-law}
\delta S= -\sum_i \Omega'^i {\delta J_i}+\sum_p \Phi'^p\delta q_p\,,
\end{align}
where
\be
{\Omega'}^i=\dfrac{\partial \Omega^i}{\partial T_H}\bigg|_{T_H=0}\,,\qquad {\Phi'}^p=\dfrac{\partial \Phi^p}{\partial T_H}\bigg|_{T_H=0}\,.
\ee
Eq.\eqref{limiting-first-law} reduces to the NHEG entropy perturbation law \eqref{entrop-pert-law}, if we show that $k^i=-\dfrac{1}{2\pi}\dfrac{\partial \Omega^i}{\partial T_H} , e^p=\dfrac{1}{2\pi}\dfrac{\partial \Phi^p}{\partial T_H}$. That is what we will do next.
\subsection{Interpretation of $k^i,\ e^p$}
To relate $\Omega'^i$ and $\Phi'^p$  (which are constructed from thermodynamic chemical potential of black holes in the extremal limit) to the $k^i$ and $e^p$ which are parameters appearing in the NHEG, after taking the near horizon limit, we need to make a connection between process of taking the \emph{near extremal} limit and the \emph{near horizon} limit performed in section \ref{sec6.1-NH-limit}. Explicitly, we need to relate spatial derivatives of $\omega^i$ to the derivative of $\Omega^i$ (which is $\omega^i$ computed at the horizon) with respect to temperature. ($\omega^i$ are defined in \eqref{BH geometry}.) Similar arguments may also be repeated for the electric charges and the corresponding potentials.  To do so, we use  the values of the chemical potentials at inner and outer horizons and the corresponding continuity conditions.

Any function in the black hole solution (like metric components) has a spacetime and a parametric dependence. Here we choose $T_H$ and the conserved charges $J_i,q_p$ as the basis for parameter space of a generic black hole; the subspace $T_H=0$ specifies the extremal black holes.
In order to relate $\partial_\rho \omega$ and thermodynamic quantities of black hole, we use a novel symmetry of black holes pointed out in \cite{Chen:2012mh} based on ideas initiated in \cite{Larsen:1997ge}. We call it horizons permutation symmetry (see appendix C for a proof) which states that under $r_+\leftrightarrow r_-$,
\begin{align}
&\Omega^i_+ \longleftrightarrow \Omega^i_-\,,\cr
&\Phi^p_+\longleftrightarrow \Phi^p_-\,,\\
&\kappa_+ \longleftrightarrow -\kappa_-\,,\nn
\end{align}
where $\Omega^i_\pm,\ \Phi^p_\pm,\kappa_\pm$ are respectively the angular velocity, gauge field potential, and surface gravity of outer/inner horizons.
This symmetry takes a more convenient form in terms of $r_h,\epsilon$ defined in \eqref{r_h}, as
\begin{align}
 r_\pm = r_h(T_H,J,...) \pm \epsilon\,.
\end{align}
Since for small $\epsilon$ temperature is proportional to $\epsilon$, $r_h=r_h(\epsilon,J,...)$, and $r_h\rightarrow r_e$ as we take $\epsilon\rightarrow 0$. As the first step we  prove that corrections to $r_h$ as we move away from $r_e$ grow like $\epsilon^2$ in the leading order.
\begin{proof} We first note that
$r_h=(r_++r_-)/2\  \text{is symmetric under } r_+ \longleftrightarrow r_-\,, \text{while} \ 2\epsilon=r_+-r_- \ \text{is antisymmetric}$.
This in particular implies that $r_+ \longleftrightarrow r_-$ transformation is equivalent to $\epsilon \longleftrightarrow -\epsilon$ or  $T_H\longleftrightarrow -T_H$ transformation. Therefore, $r_h(\epsilon)=r_h(-\epsilon)$ and $\dfrac{\partial r_h}{\partial \epsilon}=0$ or  $r_h=r_e +\mathcal{O}(\epsilon^2)$.
\end{proof}
We should comment that in the above analysis, we started with $T_H \geq 0$ but extended the parameter space of black holes to the negative $T_H$ as well.
The point $(-T_H,J)$ describes the \textit{inner horizon} of the black hole with $(T_H,J)$ and the transformation $T_H\rightarrow -T_H$ reveals the inner horizon thermodynamics \cite{Chen:2012mh}. From the black hole geometry viewpoint, this is equivalent to moving from $r_+$ to $r_-$ and hence we have built the  connection between moving in the radial direction in spacetime  and moving in the parameter space of black holes, from which we can deduce our desired relations.

We now prove that radial derivative of $\omega^i(\rho)=g^{ij}g_{tj}$ can be related to the parametric derivative of horizon angular velocity $\Omega^i_\pm$ w.r.t temperature, i.e
\begin{align}\label{1to2}
\frac{\partial \omega^i}{\partial \rho}\bigg|_{\rho=r_e}=\pm\frac{\partial \Omega^i_\pm}{\partial \epsilon}\bigg|_{\epsilon=0}\,.
\end{align}
\begin{proof}
The $r_+\rightarrow r_-\Rightarrow \Omega^i_+\rightarrow \Omega^i_-$ symmetry, in the lowest order in $\epsilon$ yields
\begin{align}\label{symmetry}
\Omega^i_+ -2\epsilon \dfrac{\partial \Omega^i}{\partial \epsilon}=\Omega^i_-\Rightarrow \Omega^i_+-\Omega^i_-=2\epsilon \dfrac{\partial \Omega^i}{\partial \epsilon}\,,
\end{align}
where $\Omega^i$ is the (outer) horizon angular velocity $\Omega^i_+$.
On the other hand, by definition of $\Omega^i$ we have
\begin{align}\label{omegapm}
\Omega^i_\pm=\omega^i(r_\pm;J,\epsilon)\Rightarrow \Omega^i_+-\Omega^i_-=2\epsilon \dfrac{\partial \omega^i}{\partial \rho}\bigg|_{\rho=r_e}\,,
\end{align}
and hence
\begin{align}
\frac{\partial \omega^i}{\partial \rho}\bigg|_{\rho=r_e}=\frac{\partial \Omega^i}{\partial \epsilon}\bigg|_{\epsilon=0}\,.
\end{align}
Similarly one can show that
\begin{align}
\frac{\partial \Phi^{(p)}(\rho)}{\partial \rho}\bigg|_{\rho=r_e}=\frac{\partial \Phi^{(p)}}{\partial \epsilon}\bigg|_{\epsilon=0}\,.
\end{align}
\end{proof}

This is an interesting identity because ${\partial \omega}/{\partial \rho}$ is completely geometrical and concerns the change of $\omega$ by moving outside the horizon of an extremal black hole, but ${\partial \Omega}/{\partial \epsilon}$ is a quantity in the parameter space and measures the change of angular velocity by turning the temperature on, and has no geometrical meaning.

We can now compute $k^i$ in \eqref{parameters}:
\begin{align}
k^i =-\dfrac{1}{CD} \dfrac{\partial \omega}{\partial \rho}\bigg|_{\rho=r_e}%-\dfrac{1}{CD} \dfrac{\partial \Omega_H}{\partial \epsilon}\right\vert_{J}\\
=\dfrac{1}{2\pi}\dfrac{\partial \Omega^i}{\partial T_H}\bigg|_{\epsilon=0}\,,
\end{align}
where we used \eqref{temperature}. One may similarly work out $e^p$, and with these in hand \eqref{limiting-first-law} takes the form
\begin{align}
\delta S=2\pi\left(\sum_i k^i\delta J_i+\sum_p e^p \delta q_p\right)\,.
\end{align}
That is, we have obtained NHEG entropy perturbation law as the appropriate near extremal limit of the first law of black hole thermodynamics.

%-------------------------------------------------------------------------------------------------------------------------------------------------------------
\section{Concluding remarks}\label{sec-discussions}

In this work we focused on the NHEG as a well-studied and classified solution to gravity theories and worked out universal relations among the parameters defining these solutions and the corresponding conserved charges. In particular we pointed out three laws of NHEG dynamics: (1) $k^i$ and $e^p$ parameters defining the NHEG are constants. (2) We have the ``entropy law'' which relates entropy (as a Noether charge) associated with the NHEG to conserved charges angular momenta $J^i$ and the electric charges $q^p$ and the on-shell value of Lagrangian (integrated over $H$), and (3) the ``entropy perturbation law,'' which relates entropy and other Noether charges associated with a probe (probing the NHEG background) to each other.

The entropy and entropy perturbation laws, despite the similarity to laws of black hole thermodynamics do not indeed have a thermodynamical interpretation; in the NHEG case we are dealing with a system which cannot be excited (without destroying the \sltr\ isometry) \cite{First-law, Amsel:2009ev}. Among other points, we would like to stress that the entropy law does not have a correspondent in the black hole thermodynamics systems. Technically, this is due to the fact that in the Wald's derivation of the first law for black holes there are ambiguities defining the charge integrals which prevents one to draw a universal relation among the thermodynamical parameters of black holes, while such ambiguities does vanish when we consider variations of fields and the corresponding perturbations in the thermodynamical charges, as they appear in the first law  of thermodynamics.

It is worth also mentioning that the entropy and entropy perturbation laws are invariant under permutation of $N$ $U(1)$ symmetries. Under these permutations
$k^i$ and $e^p$ and the corresponding charges are rotated into each other, while $S$ and $\delta S$ are only a function invariant under these permutations.
It is interesting to explore this permutation symmetry further.

Regarding the entropy perturbation law, as we discussed $\delta S$, $\delta J_i$ and $\delta q_p$ are associated with a field configuration 
$\delta \Phi$ probing the NHEG background, given by the field configuration $\Phi_0$. As we argued, entropy perturbation law \eqref{entrop-pert-law} 
is valid for $\delta\Phi$ satisfying  equations of motion linearized around background $\Phi_0$. Moreover, $\delta\Phi$ should be such that
$\delta \mathcal{E}_a=0.$ Given the discussions in \cite{Amsel:2009ev} one may wonder if these two conditions can be satisfied. Our preliminary analysis \cite{Progress} shows the answer is positive. In answering this question one may also explore if there is any relation between these $\delta\Phi$ and the set of perturbations and boundary conditions appearing in the Kerr/CFT proposal \cite{Kerr/CFT,{Compere-higher-derivative}}. It is also desirable to understand better the connection of our derivations and the NHEG mechanics with the entropy function analysis. This is also postponed to future works.

%{One may consider variations which also change the NHEG background; e.g. consider  two ``nearby'' NHEG's, one at $(k^i, e^a)$ and the other at $(k^i+\hat\delta k^i, e^a+\hat\delta e^a)$, and view the ``difference between'' the field configurations for these solutions as $\hat\delta\Phi$ probes. To distinguish such variations from the perturbations discussed in  \ref{sec-5-entropy-variation}, we have denoted them by $\hat\delta$. In this case one may simply vary the entropy law \eqref{entropy law} to obtain $J_i\hat\delta k^i+Q_a\hat\delta e^a=\hat\delta \oint_H  \sqrt{-g}{\cal L}$. This point will be explored and discussed further in upcoming publications \cite{Progress}.}

In general, especially when we deal (extremal) black holes of non-trivial horizon topology, it is possible to have solutions with non-zero ``dipole charges''. One such example is the neutral singly rotating dipole black ring \cite{Emparan-dipole-ring}. The dipole charge in fact contributes to the energy of the system and appears both in first law or the Smarr-type relation for the dipole black ring \cite{Emparan-dipole-ring}. Following Wald's derivation for the first law one can in fact prove that in general such dipole charges should appear in the first law \cite{Copsey-Horowitz}. In principle black holes/rings with dipole charges can become extremal. For example  the five dimensional dipole black ring of \cite{Emparan-dipole-ring} can become extremal while the dipole charge is still non-zero. One may study near horizon limit of extremal dipole rings and see that they exhibit \sltr$\times$U$(1)^2$ \cite{Kunduri:2007vf} and hence they fall into our definition of the NHEG. One then expects these dipole charges to appear both in our entropy law and in the entropy perturbation law \cite{Progress}.

One may wonder if the second law of thermodynamics has a correspondent in the NHEG case. Here we make a comment on that and postpone a more thorough analysis to the future publications. Let us for simplicity consider the NHEG ansatz \eqref{NHEG metric}. One may show that the angular momentum $J_i$ is given by the Noether integration
\be
J_i\propto  \int_H F(\theta) \gamma_{ij} k^j\quad \Longrightarrow \quad k^iJ_i \propto  \int_H F(\theta) k^i \gamma_{ij} k^j\,,
\ee
where $F(\theta)$ is a positive definite function and $\gamma_{ij}$ is also a positive definite metric on the $\phi^i$ part of the NHEG geometry. Therefore, $k^i J_i$ is positive definite. Similar relation also holds for $e^pq_p$.

We also discussed a derivation of NHEG mechanics laws from near extremal black holes, this latter amount to finding a relation between spatial derivatives of black hole metric functions and the parametric derivatives of the chemical potentials (horizon angular velocities or electric potentials).  To this end we proved and used the inner-outer horizon exchange symmetry (see discussions in section \ref{sec-6-NHEGvs.EBH} and appendix \ref{appendix-inner-outer-exchange}). It is desirable to understand this symmetry better and study its further implications.

%%%%%%%%%%%%%%%%%%%%%%%%%%%%%%%%%%%%%%%%%%%%%%%%%%%%%%%%%%%%%%%%%%%%%%%%%%%%%%%%%%%%%%%%%%%%%%%%%%%%%%%%%%
\section*{Acknowledgement}

We would like to thank Bin Chen, Geoffrey Comp\`{e}re,  Monica Guica, Finn Larsen,  Saeedeh Sadeghian, Joan Sim\'{o}n, Hossein Yavartanoo, Jia-ju Zhang for fruitful discussions and/or comments on the draft. We would like to also thank the anonymous referee for very useful comments which helped us improving the presentation of our results. 
KH and AS would like to thank Mehdi Golshani for his encouragement and guidance. MMShJ would like to thank organizers of
``Open Questions in Open Universe'' workshop, held in August 2013 in Istanbul, where  preliminary results of this work was presented. We would also like to thank organizers of workshop ``Quantum Aspects of Black Holes and its Recent Progress'', held in Yerevan,  September 2013, where this work was presented.

%-------------------------------------------------------------------------------------------------------------------------------------
%-------------------------------------------------------------------------------------------------------------------------------------
\appendix

%-------------------------------------------------------------------------------------------------------------------------------------
\section{On $sl(2,\mathbb{R})$ Lie algebra }\label{sec SL2R algebra}

\sltr\ is the group of all $2\times 2$ real-valued matrices with determinant one. The sl$(2,\mathbb{R})$ Lie algebra
with generators $\xi_a,\ a=1,2,3$ is defined as
\begin{align}
[\xi_a,\xi_b]=f_{ab}^{\,\,\,\;c}\xi_c
\end{align}
where $f_{ab}^{\,\,\,\;c}$ are structure constants. In this paper we have chosen the basis in a way that the commutation relations take the form
\begin{align}\label{eq sl2r generators app}
\left[\xi_1,\xi_2\right]=\xi_1\,,\qquad \left[\xi_2,\xi_3\right]=\xi_3\,,\qquad
\left[\xi_1,\xi_3\right]&=\xi_2\,.
\end{align}
In this basis, the Killing form (metric) of the algebra is
\begin{equation}\label{Killing form}
K_{ab}=\begin{pmatrix}
0&0&1\\
0&-1&0\\
1&0&0
\end{pmatrix}
\end{equation}
and its inverse $K^{ab}=(K_{ab})^{-1}$ has the same components as itself (in the chosen basis). Metric $K_{ab}$ can be used for lowering or raising the $sl(2,\mathbb{R})$ indices, e.g. $f_{abc}=K_{cd}f_{ab}^{\;\;\;d}$. One may also show that
\be\label{f-f-sl2r}
f_{ab}^{\;\;\ c}f^{abd}=2K^{cd}\,.
\ee
One specific representation of the sl$(2,\mathbb{R})$ algebra, which also realized the \sltr\ isometry of \eqref{NHEG metric}, is given in \eqref{sl2r-generators-xi-a}.

{ \sltr\  which is a double cover of $SO(2,1)$ is also the isometry group of AdS$_2$ manifold, defined as the set of points with square distance $-1$ from the origin of a flat $1+2$ dimensional Minkowski space. In a suitable coordinate system in which the metric is \eqref{Killing form}, this condition is explicitly
\be\label{ads-2-embedding}
n^an_a=K^{ab}n_an_b=-1\,,
\ee
where $x_a=n_a$ are the position of points of AdS$_2$ in the embedding space. coordinates. A solution for $n_a$, parametrized with two parameters $t,r$ is
\be\label{n-a-r-t-append}
n_1=-r\,,\qquad n_2=-tr\,,\qquad n_3=-\frac{t^2r^2-1}{2r}\,,
\ee
then the induced metric on the AdS$_2$ surface is
\begin{align}
ds^2=-r^2dt^2+\dfrac{dr^2}{r^2}
\end{align}
which is the metric of \ads{2} in Poincar\'{e} patch.
}
The $n_a,\ a=1,2,3$ form a vector representation under \sltr\ and hence,
\begin{equation}\label{eq identity xi n}
\delta_{\xi_a}n_b=f_{ab}^{\;\;\ c}n_c\,,\qquad \delta_{\xi_a}(n_bn^b)=0\,,
\end{equation}
where $\delta_{\xi_a}n_b$ is the Lie derivative of the vector $n_b$. Using the explicit form of \eqref{sl2r-generators-xi-a} and \eqref{n-a-r-t}
one may show that
\begin{equation}\label{eq identity 3}
n^a\delta_{\xi_a} n_b=0\,,\qquad \delta_{\xi_a}n_b=\xi_a^t\xi_b^r-\xi_a^r\xi_b^t\,.
\end{equation}

The above relations also show that the constant $r,t$ part of the NHEG metric \eqref{NHEG metric},  the codimension two surface $H$, is an \sltr\ invariant space, i.e. its metric and volume form do not depend on which constant $r,t$ the surface $H$ is defined.

\mydef The \emph{binormal} tensor of the $SL(2,\mathbb{R})$ invariant surfaces $H$ is defined as:
\begin{equation}\label{def-binormal}
\epsilon _{\mu\nu}\equiv\xi^a_\mu \nabla_\nu n_a\,.
\end{equation}
In the basis \eqref{sl2r-generators-xi-a} and coordinate \eqref{n-a-r-t}, this tensor can be calculated as follows:
\begin{align}\nn
\epsilon_{\mu\nu}=\xi_{a\:\mu}\partial_\nu n^a=K^{ab}\xi_{a\:\mu}\partial_\nu n_b
=K^{ab}\xi_{a\:\mu}(\delta^t_\nu\xi_b^r-\delta^r_\nu\xi_b^t)\,,\nn
\end{align}
where in the last equality we used
$\partial_r n_a =-\xi_a^t,\ \ \partial_t n_a =\xi_a^r$. Explicit computation for $\mu=r,t$ and with metric \eqref{NHEG metric} yields
\begin{align}
K^{ab}\xi_{a\: r}\xi_b^r=K^{ab}\xi_{a\: t}\xi_b^t=-\Gamma\,,
\end{align}
and zero for the other components. The final result is that
\begin{align}
\epsilon_{\mu \nu}=\left\lbrace
\begin{array}{l}
\epsilon_{tr}=-\epsilon_{rt}=\Gamma\,,\\
0 \hspace*{1cm}\text{other components}\,,
\end{array}\right.
\end{align}
or as a 2-form
\begin{align}
\epsilon=\Gamma dt\wedge dr=\dfrac{1}{\sqrt{-g^{tt}gr^{rr}}} dt\wedge dr\,.
\end{align}
One can also readily show that
\begin{equation}
\epsilon^2\equiv \epsilon _{\mu\nu}\epsilon^{\mu \nu}=-2
\end{equation}
\subsection{\ads{2} in global coordinates, another example}
As another example, let us consider NHEG in the \emph{global} coordinate for \ads{2}:
\begin{align}\label{NHEG global metric}
ds^2=\Gamma\left[-(1+r^2)dt^2+\dfrac{dr^2}{1+r^2}+\sum_{\alpha,\beta=1}^{d-n-3} {\Theta}_{\alpha \beta}d\theta^\alpha d\theta^\beta+\sum_{i,j=1}^n{\gamma}_{ij}(d\varphi^i+k^irdt)(d\varphi^i+k^jrdt)\right]
\end{align}
where $\Gamma,\Theta_{\alpha \beta},\gamma_{ij}$ are some functions of $\theta^\alpha$, specified by the equations of motion. Associated with this coordinate system, the sl$(2,\mathbb{R})$ Killing vector fields are given as
\begin{align}
\xi_1&=\partial_t\,,\nonumber\\
\xi_2&=\sin t\frac{r}{\sqrt{1+r^2}}\partial_t- \cos t{\sqrt{1+r^2}}\partial_r+\sin t \sum_{i=1}^n\frac{k^i}{\sqrt{1+r^2}}\partial_{\varphi^i}\,, \\
\xi_3&=-\cos t\frac{r}{\sqrt{1+r^2}}\partial_t- \sin t{\sqrt{1+r^2}}\partial_r-\cos t \sum_{i=1}^n \frac{k^i}{\sqrt{1+r^2}}\partial_{\varphi^i}\,.\nn
\end{align}
In this basis the sl$(2,\mathbb{R})$ commutation relations and metric are
\begin{align}\label{eq sl2r generators global}
\left[\xi_1,\xi_2\right]=-\xi_3\,,\qquad \left[\xi_3,\xi_1\right]=-\xi_2\,,\qquad\left[\xi_2,\xi_3\right]=\xi_1\,,
\end{align}
\begin{equation}\label{sltr-metric}
K_{ab}=\begin{pmatrix}
1&0&0\\
0&-1&0\\
0&0&-1
\end{pmatrix}\,.
\end{equation}

The solution to \eqref{ads-2-embedding} which also satisfies \eqref{eq identity xi n} is now given as
\begin{align}
n_1=-r\,,\qquad n_2=-\sqrt{1+r^2}\sin t\,,\qquad n_3=\sqrt{1+r^2}\cos t\,.
\end{align}
It can be checked that relations $\partial_r n_a =-\xi_a^t$ and $\partial_t n_a =\xi_a^r$ also hold in the global coordinate and hence \eqref{eq identity 3} is still true. Using the same discussion as above one can show that using the definition \eqref{def-binormal} leads to the same result for the binormal tensor
\begin{align}
\epsilon_{\mu\nu}=\dfrac{1}{\sqrt{-g^{tt}gr^{rr}}} dt\wedge dr\,.
\end{align}

%-------------------------------------------------------------------------------------------------------------------------------------
\section{Symmetries and conserved charges}\label{sec symmetry and charges}
\emph{Symmetry} is a transformation which maps a set of solutions of equations of motion (with appropriate boundary conditions) to themselves and hence leaves the action invariant, or equivalently, changes the Lagrangian up to a total divergence. The symmetries could be local (gauge) or global and both of these have been argued to be a basis for deriving constants of motion or conserved charges, see \cite{Symmetry-charges} and references therein for a historical review.
Here we will be mainly concerned with symmetries associated with spacetime coordinate transformations and diffeomorphisms and will follow Wald's papers
\cite{Wald:1993nt,Iyer:1994ys,{Wald-review}}.

Consider a diffeomorphism invariant theory with a Lagrangian density $\mathcal{L}$ and the corresponding action in $d$-dimensional space-time
\begin{equation}
I[\phi]=\int \mathrm{d} ^d x \sqrt{-g} \mathcal{L}(\Phi ; x^ \mu)
\end{equation}
in which $\Phi$ denotes all of dynamical fields of the system and each of them will be denoted by $\Phi ^ i$. Associated with any infinitesimal diffeomorphism as a symmetry of the theory, one can find a Noether current and the corresponding Noether charge.  Following \cite{Iyer:1994ys} we take the Lagrangian $\mathbf{L}$  to be a top form, a $d$-form equal to $\sqrt{-g}\mathcal{L}\mathbf{\epsilon}_d$ with $\mathbf{\epsilon}_d$ being the Levi-Civita tensor, and  generator of  diffeomorphism symmetry to be a 1-form $\xi$. Variation of Lagrangian under the diffeomorphism is \cite{Wald'90}
\begin{equation}\label{eq lagrangian deviation}
\delta _\xi \mathbf{L}=\mathbf{ E}_i \delta _\xi\Phi ^i+\mathrm{d}\mathbf{\Theta} (\Phi , \delta_\xi\Phi)\,,
\end{equation}
where $\mathbf{E}_i=0$ is the e.o.m for $\Phi^i$. The $(d-1)$-form $\mathbf{\Theta}$ is the surface term generated by the variation.% and is (Hodge) dual to  1-form $\mathbf{\Theta} _\mu \mathrm{d} x^\mu$.
%\begin{equation}
%\mathbf{\Theta} ^ \mu = \frac{\partial \mathcal{L}}{\partial (\partial _\mu \Phi ^i)}\delta_\xi \Phi ^i
%\end{equation}

According to the identity $\delta _\xi \mathbf{L}= \xi \! \cdot \! \mathrm{d} \mathbf{L} +\mathrm{d} (\xi \! \cdot \! \mathbf{L})$ and noting that $\mathrm{d} \mathbf{L}=0$, we can replace the LHS of \eqref{eq lagrangian deviation}:
\begin{equation}
\mathrm{d}\mathbf{\Theta}(\Phi , \delta_\xi\Phi)-\mathrm{d} (\xi \! \cdot \! \mathbf{L}) =-\mathbf{E}_i \delta _\xi\Phi ^i
\end{equation}
Now, we can associate a Noether $(d-1)$-form  current $\mathbf{J}$ as:
\begin{equation}\label{current}
{\mathbf{J} \equiv \mathbf{\Theta}(\Phi , \delta_\xi\Phi)-\xi \! \cdot \! \mathbf{L}}
\end{equation}
Therfore $\mathrm{d}\mathbf{ J}=-\mathbf{E}_i \delta _\epsilon\Phi ^i$ so that $\mathrm{d} \mathbf{J}=0$ whenever e.o.m is satisfied and according to the Poincar\'e's lemma, since $\mathbf{J}$ is closed, it would be exact and can be written as:
\begin{equation}\label{Q-form}
{\mathbf{J}=\mathrm{d} \mathrm{\mathbf{Q}}}
\end{equation}
where $\mathrm{\mathbf{Q}}$ is a $(d-2)$-form, the \emph{ Noether charge density}.
%------------------------------------------------------------------------------------------------------------------------------------------------------------------------------------------------

\subsection{Ambiguities}\label{sec ambiguities}

It has been shown \cite{Iyer:1994ys,Wald-review} that the $(d-1)$-form $\mathbf{J}$ in \eqref{current} has twofold ambiguities. One ambiguity comes from freedom of the definition of  Lagrangian of the theory up to an exact $d$-form:
\begin{equation}
\mathcal{L} \to \mathcal{L}+d \mu\,,
\end{equation}
which leads to $\mathbf{J} \to \mathbf{J} +\delta_\xi \mu$. The other ambiguity comes from the freedom in specifying $\mathbf{J}$ itself (for a given  Lagrangian) up to an exact $(d-1)$-form $d {Y}(\Phi, \delta \Phi)$. Therefore, the Noether current $\mathbf{J}$ is defined up to the following ambiguities
\begin{equation}\label{eq J ambiguities}
\mathbf{J} \to \mathbf{J} + d (\xi \cdot \mu) +d {Y}(\Phi, \delta \Phi)\,,
\end{equation}
where the $(d-2)$-form ${Y}(\Phi, \delta \Phi)$ is linear in $\delta_\xi \Phi$ and  we used the identity $\delta_\xi \mu=\xi \cdot d \mu+d (\xi \cdot \mu)$. When we want to find the Noether charge, in addition to these ambiguities there is another one which is the freedom of choosing $\mathbf{Q}$ up to an exact $(d-2)$-form $d {Z}(\Phi,\xi)$ where ${Z}$ is linear in $\xi$. So accumulating all of the ambiguities, we have the freedom of choosing the Noether charge density as:
\begin{equation}
\mathbf{Q} \to \mathbf{Q} +\xi \cdot \mu +{Y}+d{Z}\,,
\end{equation}
and hence  the Noether charge density $\mathbf{Q}$ is not unique and its most general is  \cite{Iyer:1994ys}
\begin{equation}\label{eq decomposition}
\mathbf{Q}=W_\mu(\Phi)\xi^\mu+\mathbf{E}^{\mu \nu}(\Phi)\nabla_{[\mu}\xi _{\nu ]}+Y(\Phi, \delta_\xi \Phi)+d Z(\Phi , \xi),
\end{equation}
where $W_\mu$ and $\mathbf{E}^{\mu \nu}$ and $Y$ and $Z$ are covariant quantities which are locally constructed from fields and their derivatives, $Y$ is linear in $\delta _\xi \Phi$, $Z$ is linear in $\xi$ and,
\be\label{E-four-index}
(\mathbf{E}^{\mu\nu})_{\alpha_3 \dots \alpha_d}=-{E}^{\alpha \beta \mu \nu}\epsilon_{\alpha \beta\alpha_3 \dots \alpha_d }\,,\qquad {E}^{ \mu \nu \alpha \beta }\equiv \frac{\delta \mathcal{L}}{\delta R_{\mu \nu \alpha \beta}}\,.
\ee
In order to fix/remove these ambiguities, we need some physical reasoning and/or reference point for defining the charges (like requesting to coincide with the ADM charges etc.)

%-------------------------------------------------------------------------------------------------------------------------------------------------------------------------------------------------------------------------

\subsection{Iyer-Wald entropy}\label{sec Iyer Wald}
Iyer-Wald entropy \cite{Wald:1993nt,Iyer:1994ys} for a generic stationary  black hole with \emph{bifurcate horizon} is defined as:
\begin{equation}
{\frac{S}{2 \pi}  \equiv -\int_\mathcal{H} \text{Vol}(\mathcal{H}) \frac{\delta \mathcal{L}}{\delta R_{\mu \nu \alpha \beta}}\epsilon_{\mu \nu}\epsilon_{\alpha \beta}}
\end{equation}
where $\epsilon_{\alpha \beta}=n_{[ \alpha}\xi_{ \beta]}$ is the \emph{binormal} for the $d-2$-dimensional horizon surface $\mathcal{H}$ and the vectors $\xi _\mu$ and $n_\mu$ are normals to the bifurcate horizon null surface which on the horizon satisfy the relations
\begin{equation}\label{eq normals}
n\!\cdot \! n=0\,,\qquad \xi \!\cdot \! \xi =0\,,\qquad n \!\cdot \! \xi =-1
\end{equation}
and according to them, the binormal satisfies $\epsilon ^2=-2$.

\section{Computation of symplectic form}\label{appendix symplectic form}
{Here we present details of computation of the symplectic form appearing in the LHS of \eqref{Omega}. As mentioned \cite{Iyer:1994ys}, the symplectic current $\boldsymbol{\omega}=0$ for $\delta_\xi\Phi=0$. This is true for the Killing vectors of NHEG, except for $\xi_3$ when acting on gauge fields where there is a residual gauge transformation. To compute the effects of this residual gauge transformation, we start with the definition of $\boldsymbol{\omega}$ 
\begin{align}
\boldsymbol{\omega}(\Phi,\delta\Phi,\delta_\xi\Phi)= \delta \mathbf{\Theta}(\delta_\xi\phi)-\delta_\xi \mathbf{\Theta}(\delta\phi)\,.
\end{align}
We discussed the form of $\mathbf{\Theta}$, or its Hodge dual vector field $\Theta^\mu$, for gauge fields in \eqref{gauge-field-contrib.}:
\begin{align*}
\Theta^\mu(\delta A_\alpha)=\dfrac{\partial \mathcal{L}}{\partial F_{\mu\nu}}\delta A_\nu,%\hspace{1cm}\Theta(\delta A_\alpha)=\star\,\Theta^\mu(\delta A_\alpha)
\end{align*}
so
\begin{align}
\delta_2\Theta^\mu(\delta_1 A_\alpha)&=\delta_2(\dfrac{\partial \mathcal{L}}{\partial F_{\mu\nu}}\delta_1 A_\nu)\\
&=\delta_2(\dfrac{\partial \mathcal{L}}{\partial F_{\mu\nu}})\;\delta_1 A_\nu+\dfrac{\partial \mathcal{L}}{\partial F_{\mu\nu}}\;\delta_2\delta_1 A_\nu\,.
\end{align}
Assuming that $\delta_1\delta_2=\delta_2\delta_1$ (which is true for $\delta,\delta_\xi$)
\begin{align}
\omega^\mu(\Phi,\delta_1\Phi,\delta_2\Phi)=\delta_2(\dfrac{\partial L}{\partial F_{\mu\nu}})\delta_1 A_\nu-\delta_1(\dfrac{\partial L}{\partial F_{\mu\nu}})\delta_2 A_\nu\,,
\end{align}
where $\omega^\mu$ is the vector Hodge dual to the $(d-1)$-form symplectic current $\boldsymbol{\omega}$.
The nonvanishing part of $\omega$ is hence %due to nonvanishing of $\delta_{\xi_3}A_\mu$ is
\begin{align}
\omega^\mu(\Phi,\delta\Phi,\delta_{\xi_3}\Phi) &=\delta(\dfrac{\partial L}{\partial F_{\mu\nu}})\delta_{\xi_3} A_\nu-\delta_{\xi_3}(\dfrac{\partial L}{\partial F_{\mu\nu}})\delta A_\nu\\
&=\delta(\dfrac{\partial L}{\partial F_{\mu\nu}})\delta_{\xi_3} A_\nu\,.
\end{align}
The second term on the right hand side is zero since $\xi_3$ is a symmetry of Lagrangian and $F_{\mu\nu}$.
Next, recall from \eqref{delta3} that
\begin{align*}
\delta_{\xi_3} A_\nu=(0,-\dfrac{e}{r^2},0,0)=\nabla_\nu \Lambda ,\hspace*{1cm}\Lambda=\dfrac{e}{r}
\end{align*}
therefore
\begin{align}
\nonumber\omega^\mu(\Phi,\delta\Phi,\delta_{\xi_3}\Phi) &=\delta(\dfrac{\partial L}{\partial F_{\mu\nu}})\nabla_\nu \Lambda\\
\nonumber &=\nabla_\nu\Big(\Lambda\delta(\dfrac{\partial L}{\partial F_{\mu\nu}})\Big)-\Lambda\nabla_\nu\delta(\dfrac{\partial L}{\partial F_{\mu\nu}})\\
&=\nabla_\nu\Big(\Lambda\delta(\dfrac{\partial L}{\partial F_{\mu\nu}})\Big)
\end{align}
where we have used the linearized equation of motion for the gauge field perturbations $\delta A_\mu$. Therefore, we obtain
\begin{align}
\Omega(\Phi,\delta\Phi,\delta_{\xi_3}\Phi) &=\int_\Sigma d\Sigma_\mu\omega^\mu(\Phi,\delta\Phi,\delta_{\xi_3}\Phi) = \int_\Sigma d\Sigma_\mu \nabla_\nu\Big(\Lambda\delta(\dfrac{\partial L}{\partial F_{\mu\nu}})\Big)\\
&=\oint_{\partial \Sigma}d\Sigma_{\mu\nu}\;\Lambda\;\delta(\dfrac{\partial L}{\partial F_{\mu\nu}})\,,
\end{align}
where $\Sigma$ is a constant time slice bounded between $r=r_H$ and $r=\infty$.
 $\Omega$ will hence have a term at infinity and a term on $H$. The term at infinity does not contribute since $\Lambda=\dfrac{e}{r}$ vanishes there (in fact $r=\infty$ boundary was chosen precisely for this reason). So, the only contribution is
 \begin{align}\label{Lambda-infinity}
\Omega (\Phi,\delta\Phi,\delta_{\xi_3}\Phi)&= \oint_{H}d\Sigma_{\mu\nu}\Big(\Lambda\delta(\dfrac{\partial L}{\partial F_{\mu\nu}})\Big)\\
&=\dfrac{e}{r_H}\delta q
\end{align}
where $\delta q=\oint_{H}d\Sigma_{\mu\nu}\delta(\frac{\partial L}{\partial F_{\mu\nu}})$.
Noting that $\zeta_H=n_H^a\xi_a-k^im_i$, and that $\Omega$ is linear in $\delta_{\zeta_H}A=n_H^a\delta_{\xi_a}A-k^i\delta_{m_i}A$, we obtain
\begin{align}
\Omega(\Phi,\delta\Phi,\delta_{\zeta_H}\Phi)=n_H^3\Omega(\Phi,\delta\Phi,\delta_{\xi_3}\Phi)=n_H^3\dfrac{e^p}{r_H}\delta q_p=-e^p\delta q_p
\end{align}}

\section{Inner/outer horizons permutation symmetry}\label{appendix-inner-outer-exchange}

In this appendix we state and prove the permutation symmetry of black hole horizons.
%Consider a black hole with horizons at $r=r_i$.
Permutation symmetry states that:\footnote{We thank Bin Chen and Jia-ju Zhang for correspondence on this point.}
\bc
\textit{Let $\{r_i\}$ denote the position of horizons of a given black hole, a permutation in black hole parameters of the form $r_i\rightarrow r_{\sigma_i}$, has the following effect on black hole horizon chemical potentials:
\begin{align}
\Omega_i\xrightarrow{r_i \rightarrow r_{\sigma_i}}\Omega_{\sigma_i}
\end{align}
}
\ec
\begin{proof}
we assume that $\Delta$ is an analytic function of $r$, then $\Delta=\sum_{m=0}^n c_mr^m$ which has $n$ roots $\{r_i\}$ and $n$ constants $c_m$,
\begin{align}\label{system}
\Delta(r_i;\lbrace c_m\rbrace)=0\,,\qquad i=1,2,\cdots, n\,.
\end{align}
If we consider $c_m$'s as unknowns and $r_i$ as given parameters, the above is a system of linear equations which can be uniquely solved to write $c_m$'s in terms of $r_m$'s which results in $c_m=c_m(r_1,r_2,...)$. Now a transformation of the form $(r_i)\rightarrow (r_{\sigma_i})$ where $\sigma$ is a permutation function of $1,2,...,n$, does not change the set of equations and as a result, the solutions $c_m=c_m(r_1,r_2,...)$ are still solutions, and from the fact that the solution is unique, this means that $c_m(r_{\sigma(i)})=c_m(r_i)$. Therefore,
\begin{align}
\Omega_i=\omega(r=r_i; \lbrace c_m\rbrace)\xrightarrow{r_i\rightarrow r_{\sigma(i)}} \:\omega(r=r_{\sigma(i)}; \lbrace c_m\rbrace)=\Omega_{\sigma(i)}\,.
\end{align}
Although $\omega$ can in principle depend on other parameters of black holes, than $c_i$, e.g $d_1,d_2, ...$ this dependence is not relevant to our argument because the transformation $r_i\rightarrow r_{\sigma(i)}$ does not change $d_i$. The reason is that we assume the system of equations \eqref{system} has a unique solution, and so $r_i$ is completely determined by $c_m$'s and does not depend on $d_m$'s and changing (permuting) $r_i$'s does not affect $d_m$'s and our argument still holds.\end{proof}

 %%%%%%%%%%%%%%%%%%%%%%%%%%%%%%%%%%%%%%%
\bibliographystyle{plain}

%\endgroup

\end{document}